\documentclass[notitlepage,nofootinbib,preprintnumbers,amssymb,superscriptaddress]{revtex4-1}
\usepackage{amsfonts,amssymb,amsmath,graphicx,color,bm}
\definecolor{ultramarine}{rgb}{0.07, 0.04, 0.56}
\definecolor{cadmiumgreen}{rgb}{0.0, 0.42, 0.24}
\definecolor{indigo(dye)}{rgb}{0.0, 0.25, 0.42}
\usepackage[linktocpage=true]{hyperref}
\hypersetup{
colorlinks=true,
citecolor=red, 
linkcolor=red, 
urlcolor=indigo(dye),
}

\newcommand{\be}{\begin{equation}}  
\newcommand{\ee}{\end{equation}}
\newcommand{\bem}{\begin{bmatrix}}
\newcommand{\eem}{\end{bmatrix}}

\allowdisplaybreaks

\begin{document}

\title{
Black holes with a nonconstant  kinetic term in degenerate higher-order scalar tensor theories
}

\author{Masato Minamitsuji}
\affiliation{Centro de Astrof\'{\i}sica e Gravita\c c\~ao  - CENTRA, Departamento de F\'{\i}sica, Instituto Superior T\'ecnico - IST,
Universidade de Lisboa - UL, Av. Rovisco Pais 1, 1049-001 Lisboa, Portugal}

\author{James Edholm}
\affiliation{
Physics Department, Lancaster University, Lancaster, LA1 4YW
}

\begin{abstract}
We investigate static and spherically symmetric black hole (BH) solutions in shift-symmetric quadratic-order degenerate higher-order scalar-tensor (DHOST) theories. We allow a nonconstant kinetic term $X=g^{\mu\nu} \partial_\mu\phi\partial_\nu \phi$ for the scalar field $\phi$ and assume that $\phi$ is, like the spacetime, a pure function of the radial coordinate $r$, namely $\phi=\phi(r)$. First, we find analytic static and spherically symmetric vacuum solutions in the so-called {\it Class Ia} DHOST theories, which include the quartic Horndeski theories as a subclass. We consider several explicit models in this class and apply our scheme to find the exact vacuum BH solutions. BH solutions obtained in our analysis are neither Schwarzschild or Schwarzschild (anti-) de Sitter. We show that a part of the BH solutions obtained in our analysis are free of ghost and Laplacian instabilities and are also mode stable against the odd-parity perturbations. Finally, we argue the case that the scalar field has a linear time dependence $\phi=qt+\psi (r)$ and show several simple examples of nontrivial BH solutions with a nonconstant kinetic term obtained analytically and numerically.
\end{abstract}

\maketitle

\section{Introduction}
\label{sec1}

Scalar-tensor theories have provided the unified mathematical description of modified gravity theories \cite{Berti:2015itd}.
In classic scalar-tensor theories,
whose Lagrangian density depends on the metric $g_{\mu\nu}$, the scalar field $\phi$, and its first-order derivative $\phi_\mu:=\nabla_\mu\phi$,
${\cal L}={\cal L} (g_{\mu\nu},\phi,\phi_\mu)$,
the Euler-Lagrange (EL) equations are given by the second-order differential equations.
However, 
the Lagrangian density of modern 
scalar-tensor theories
may also contain the second-order derivatives of 
the scalar field $\phi_{\mu\nu}:=\nabla_\mu\nabla_\nu \phi$.
Although generically
the EL equations in such theories 
contain higher derivative terms,
the appearance of Ostrogradsky ghosts~\cite{Woodard:2015zca}
can be avoided using certain degeneracy conditions~\cite{Motohashi:2016ftl}.
Degenerate higher-order scalar-tensor (DHOST) theories~\cite{Langlois:2015cwa,Crisostomi:2016czh,Achour:2016rkg,BenAchour:2016fzp,Crisostomi:2017lbg,Langlois:2017dyl}
(see also \cite{Langlois:2018dxi,Kobayashi:2019hrl})
provide the most general framework of scalar-tensor theories
which are free from Ostrogradsky instabilities \cite{Woodard:2015zca},
and hence the system contains only three degrees of freedom (DOFs),
namely, two tensorial and one scalar polarizations
(See \S~\ref{Dhostsec} for details).
Applications of DHOST theories to cosmological and astrophysical problems 
have been investigated in Refs. \cite{Langlois:2017mxy,Langlois:2017dyl,Dima:2017pwp,Creminelli:2018xsv,Saltas:2019ius}.

The application of modern scalar-tensor theories to BH physics has attracted great interest.
Besides the Schwarzschild or Kerr solutions in General Relativity (GR) with or without a constant scalar field
\cite{Motohashi:2018wdq,Motohashi:2019sen}, 
i.e. GR BH solutions, 
they also allow BH solutions which are absent in GR.
A typical nontrivial BH solution
is the \textit{stealth Schwarzschild} solution~\cite{Babichev:2013cya,Kobayashi:2014eva} 
obtained in shift-symmetric Horndeski theories
with the assumptions of a linearly time-dependent scalar field $\phi=qt+\psi (r)$
and a constant kinetic term $X={\rm const}$,
where $t$ and $r$ are the time and radial coordinates of the static and spherically symmetric spacetime
and $X:= g^{\mu\nu}\phi_\mu \phi_\nu$ represents the canonical kinetic term of the scalar field.
In stealth solutions, 
the spacetime geometry is completely independent of the coupling functions and the profiles of the scalar field.
Stealth Schwarzschild solutions have also been found in  shift-symmetric DHOST theories
in Refs. \cite{BenAchour:2018dap,Motohashi:2019sen,Takahashi:2019oxz,Minamitsuji:2019shy}.
Another nontrivial solution 
is the Schwarzschild-(anti-) de Sitter solution with the same properties of the scalar field 
\cite{Babichev:2013cya,Kobayashi:2014eva,Babichev:2017lmw,BenAchour:2018dap,Motohashi:2019sen,Takahashi:2019oxz,Minamitsuji:2019shy}.
Kerr-(anti-) de Sitter solutions with a constant $X$
were obtained in Ref. \cite{Charmousis:2019vnf,Charmousis:2019fre}.
It was recently argued, 
however,
that stealth or Schwarzschild-(anti-) de Sitter  BH solutions with the constant kinetic term suffer issues of the strong coupling
in the sector of the scalar perturbations \cite{deRham:2019gha}.
Neutron star (NS) solutions were also explicitly constructed in Ref. \cite{Kobayashi:2018xvr}.
In this paper,
we will investigate BH solutions with a nonconstant $X$
in quadratic DHOST theories.

While BH solutions with a constant $X$ can be obtained without explicitly specifying
the functions which determine the coupling of the scalar field,
those with a nonconstant $X$ will require explicit choices of these coupling functions. 
Static and spherically symmetric BH solutions obtained without the assumption of a nonconstant kinetic term 
typically have the metrics of 
neither Schwarzschild- nor Schwarzschild-(anti-) de Sitter spacetimes.
In the context of the shift-symmetric Horndeski theory,
the first exact static and spherically symmetric solutions with $\phi=\phi(r)$ and a nonconstant $X$ 
were asymptotically locally anti-de Sitter \cite{Rinaldi:2012vy,Anabalon:2013oea,Minamitsuji:2013ura}.  
An asymptotically flat solution
with the static scalar field $\phi=\phi(r)$
was also constructed 
in a class of the shift-symmetric Horndeski theory 
with a linear coupling to the Gauss-Bonnet term~\cite{Sotiriou:2013qea,Sotiriou:2014pfa}.
However, in these solutions
the Noether current $J^\mu$ does not  vanish at the event horizon 
and the norm $J_\mu J^\mu$ blows up there~\cite{Babichev:2017guv}.
The static, spherically symmetric, and asymptotically flat BH solutions 
with both the ansatz of the scalar field $\phi=\phi(r)$ and $\phi=qt+\psi (r)$,
which were different from the Schwarzschild solution,
were obtained 
in the shift-symmetric 
Horndeski and GLPV theories
with vanishing Noether current $J^\mu=0$
under the assumption
that $X$ is a nontrivial function of $r$~\cite{Babichev:2017guv}.

Recently, 
in Ref.~\cite{BenAchour:2019fdf}
static and spherically symmetric 
BH solutions with a nonconstant $X=X(r)$ in the quadratic DHOST
theories
have been explored 
by the disformal transformation of the known BH solutions. 
In this paper, 
instead, 
we will focus on analytic BH solutions with a nonconstant $X$
by the direct integration of the equations of motion.
Our approach will follow that developed in previous work
to solve the problems
in static and spherically symmetric systems \cite{Kobayashi:2018xvr,Takahashi:2019oxz}.
Although the equations of motion
are given by higher-order differential equations
with respect to the radial coordinate $r$,
because of the properties of the underlying theories,
these higher-order differential equations are degenerate
and their combination provides a constraint relation.
With this constraint relation, 
the system can reduce to that of the second-order differential equations.

The paper is constructed as follows:
In \S~\ref{Dhostsec}, we give an overview of the quadratic DHOST theories.
In \S~\ref{sec2}, 
we present the strategy to find the static and spherically symmetric vacuum solutions
with the scalar field $\phi=\phi(r)$ in the shift-symmetric Class Ia quadratic DHOST theories.
In \S~\ref{sec3},
we present the explicit models and BH solutions with a nonconstant $X$,
and investigate the odd-mode stability of these solutions.
In \S~\ref{sec5},
we argue the case of a linearly time dependent scalar field $\phi=qt+\psi (r)$,
and provide several models giving rise to notrivial BH solutions analytically and numerically.
The last \S~\ref{sec6}
is  devoted to a brief summary and conclusion.

\section{The DHOST Theories}
\label{Dhostsec}

In the context of analytical mechanics,
a simple example of degenerate higher-derivative system is given by  \cite{Langlois:2015cwa,Langlois:2018dxi,Kobayashi:2019hrl}
\begin{eqnarray}
L= \frac{a_1}{2}\ddot{x}^2
   + a_2 \ddot{x} \dot{y}
    +\frac{a_3}{2} \dot{y}^2
   + \frac{1}{2}\dot{x}^2
+\cdots,
\end{eqnarray}
where
`dot' means a derivative with respect to time;
the variables $x$ and $y$ are the straightforward analogs
of the scalar field $\phi$ and the metric $g_{\mu\nu}$
in the scalar-tensor theories, respectively;
$a_i$ ($i=1,2,3$) are assumed to be constant for simplicity.
The EL equations for $x$ and $y$ are given by 
$a_1\ddddot{x}+a_2 \dddot{y} +\cdots=0$
and 
$a_2\dddot{x}+a_3 \ddot{y}+\cdots=0$,
where dots denote the terms composed of the lower-order derivatives. 
Eliminating $y$ with the new variable $z= a_2\dot{x}+ a_3 y$,
we obtain the two independent equations
$\left(
a_2^2-a_1 a_3
\right)
 \ddddot{x}
+
\left(
\cdots
\right) \ddot{x}
+\cdots
=0$
and 
$\ddot{z}
+\cdots =0$.
Generically,
in order to determine evolution of the system, 
we need the six initial data of $x$, $\dot{x}$, $\ddot{x}$, $\dddot{x}$, $z$, and $\dot{z}$
and hence there are 2$+$1 DOFs,
where the extra one is known as an Ostrogradsky ghost \cite{Woodard:2015zca}.
However, by imposing the additional condition $a_2^2-a_1a_3=0$,
the system reduces to that of the two second-order differential equations
$\ddot{x}+\cdots=0$
and 
$\ddot{z}+\cdots=0$,
and contains 2 physical DOFs,
namely, the Ostrogradsky ghost is eliminated.
The condition like $a_2^2-a_1 a_3=0$ is called the {\it degeneracy condition}.
A degenerate theory
corresponds to a class of higher-derivative theories
where the EL equations can reduce to the second-order systems after a suitable redefinition of variables
\cite{Langlois:2015cwa,Motohashi:2016ftl,Klein:2016aiq,Motohashi:2017eya,Motohashi:2018pxg}.

DHOST theories correspond 
to the extension of the degenerate theories in analytical mechanics 
to scalar-tensor theories~\cite{Langlois:2015cwa,Achour:2016rkg,BenAchour:2016fzp,Crisostomi:2017lbg,Langlois:2017dyl}. 
The construction of DHOST theories starts
from the the most general covariant scalar-tensor Lagrangian density
which contains up to second-order covariant derivatives of the scalar field $\phi_{\mu\nu}$,
${\cal L}={\cal L} \left(g_{\mu\nu},\phi,\phi_\mu, \phi_{\mu\nu}\right)$,
and then expands it in terms of $\phi_{\mu\nu}$ as
\begin{eqnarray}
\label{full}
\frac{{\cal L}}{\sqrt{-g}}
&=&
   F_0(\phi,X)  
- F_1(\phi,X) \Box\phi
+ F_2(\phi,X) R
+ F_3 (\phi,X) G^{\alpha\beta} \phi_{\alpha\beta}
+
C_2^{\alpha\beta\rho\sigma} \left(g_{\mu\nu},\phi, \phi_\mu\right)
\phi_{\alpha\beta}
\phi_{\rho\sigma}
+ {\cal O}
\left[
(\phi_{\rho\sigma})^3
\right],
\end{eqnarray}
where 
$R$ and $G^{\alpha\beta}$ are 
the Ricci scalar and Einstein tensor associated with the metric $g_{\mu\nu}$;
$\phi_{\mu\nu\cdots\alpha}:= \nabla_\mu \nabla_\nu\cdots \nabla_\alpha\phi$
represents the covariant derivative associated with $g_{\mu\nu}$;
$X=g^{\mu\nu}\phi_\mu \phi_\nu$ is the ordinary kinetic term;
$\Box\phi:= g^{\mu\nu} \phi_{\mu\nu}$;
and 
$C_2^{\alpha\beta\rho\sigma}$ is a general rank-4 covariant tensor
constructed from $g_{\mu\nu}$, $\phi$ and $\phi_\mu$
with certain symmetries. 
We note that in the above theory \eqref{full}
the $F_0$ and $F_1$ terms correspond to a part of 
the Horndeski theory \cite{Horndeski:1974wa,Deffayet:2009wt,Deffayet:2009mn,Deffayet:2011gz,Kobayashi:2011nu}.
Truncating Eq. \eqref{full} at the order of $\phi_{\mu\nu}^2$
and setting $F_1=F_3=0$
\footnote{We set $F_1=F_3=0$, 
as we will focus on the shift- and reflection-symmetric theories.},
we obtain
\begin{eqnarray}
 \label{qdaction}
\frac{{\cal L}}{\sqrt{-g}}
=
  F_0(\phi,X)
+ F_2(\phi,X) R 
+ \sum_{I=1}^{5} A_I(\phi,X)L_I^{(2)},
\end{eqnarray}
where
\begin{eqnarray}
&& 
L_1^{(2)} =\phi_{\mu\nu}\phi^{\mu\nu},
\quad
L_2^{(2)} =  \left(\Box\phi\right)^2,
\quad
L_3^{(2)} = \Box\phi \phi^\mu \phi_{\mu\nu}\phi^\nu, 
\quad
L_4^{(2)} = \phi^\mu \phi_{\mu\rho}\phi^{\rho\nu}\phi_\nu,
 \quad
L_5^{(2)} = \left(\phi_\mu \phi^{\mu\nu}\phi_{\nu}\right)^2, 
\end{eqnarray}
where 
$A_I(\phi,X)$ ($I=1,\cdots, 5$) are functions of $\phi$ and $X$.
Degeneracy conditions for the quadratic DHOST theory \eqref{qdaction}
were obtained in Refs~\cite{Langlois:2015cwa,Achour:2016rkg,BenAchour:2016fzp}.
Among them,
we focus on the Class Ia (also known as Class ${}^2$N-I)
DHOST theories~\cite{Langlois:2015cwa,Achour:2016rkg}
for which the degenerate conditions are given by 
\begin{eqnarray}
\label{2n1}
A_2&=&-A_1\neq -\frac{F_2}{X},
\nonumber\\
A_4&=&\frac{1}{8(F_2-XA_1)^2}
\left\{
4F_2
\left[ 
3(A_1-2F_{2X})^2-2A_3F_2\right]
-A_3X^2(16A_1F_{2X}+A_3F_2) 
\right.
\nonumber\\
&&
\left.
+
4X
\left[
3A_1A_3F_2+16A_1^2F_{2X}-16A_1F_{2X}^2-4A_1^3+2A_3F_2F_{2X}
\right]
\right\}, 
\nonumber\\
A_5&=&\frac{1}{8(F_2-XA_1)^2} (2A_1-XA_3-4F_{2X}) \left(A_1(2A_1+3XA_3-4F_{2X})-4A_3F_2\right).
\end{eqnarray}
The theory \eqref{qdaction} with Eq. \eqref{2n1}
includes the quintic-order Horndeski~\cite{Horndeski:1974wa} 
and Gleyzes-Langlois-Piazza-Vernizzi (GLPV)~\cite{Gleyzes:2014dya} theories as subclasses.
The other classes of the quadratic DHOST theories
suffer Laplacian instabilities of scalar perturbations in cosmological backgrounds 
\cite{Langlois:2017mxy,Langlois:2017dyl,Langlois:2018dxi}
and are excluded from our analysis.
The almost coincident measurements
of gravitational waves (GWs) emitted from a binary NS merger by LIGO
and its associated short gamma-ray burst~\cite{TheLIGOScientific:2017qsa,GBM:2017lvd,Monitor:2017mdv,Monitor:2017mdv,Creminelli:2017sry,Ezquiaga:2017ekz,Baker:2017hug,Sakstein:2017xjx}
suggest 
that the difference between the speed of GWs $c_{gw}$ and the speed of light $c$  
is at most of ${\cal O}(10^{-15} )$.
Theories compatible with the condition $c_g=c$ on cosmological scales
have to satisfy $A_1=A_2=0$ (see also \cite{deRham:2018red}).
Further constraints on DHOST theories
are imposed in terms of graviton decay into dark energy \cite{Creminelli:2018xsv}, which put $A_3=0$.
In DHOST theories,
if the scalar field has cosmological time-dependence,
the effects of the fifth force can be enhanced inside astrophysical bodies;
constraints in terms of stellar dynamics 
have been obtained in Refs. \cite{Kobayashi:2014ida,Dima:2017pwp,Langlois:2017dyl,Saltas:2019ius}. 

\section{Static and spherically symmetric solutions in the shift-symmetric DHOST theories}
\label{sec2}

\subsection{The shift-symmetric quadratic DHOST theores}

We focus on the shift-symmetric Class Ia quadratic DHOST theories
\eqref{qdaction} with degeneracy conditions \eqref{2n1}
which are invariant under the shift of the scalar field $\phi\to \phi+c$
where $c$ is constant.
The shift symmetry can be imposed by suppressing the $\phi$ dependence from the coupling functions
in Eq. \eqref{qdaction}
\begin{eqnarray}
\label{shift}
 F_i = F_i(X), 
\qquad 
A_I = A_I(X),
\qquad
(i=0,2;~I=1, 2, 3, 4, 5).
\end{eqnarray}
We rewrite the Lagrangian density Eq. \eqref{qdaction} as
\begin{eqnarray}
\label{decomp}
{\cal L}={\cal L}_F+\sum_{i=1}^5 {\cal L}_{A_i},
\end{eqnarray}
where 
${\cal L}_F/\sqrt{-g}
:=
F_0 (X)
+F_2(X) R$,
${\cal L}_{A_1}/\sqrt{-g}
:=
 A_1(X) \phi^{\rho\sigma}\phi_{\rho\sigma}$,
${\cal L}_{A_2}/\sqrt{-g}
:=
 A_2(X) (\Box \phi)^2$,
${\cal L}_{A_3}/\sqrt{-g}
:=
 A_3(X) \Box \phi 
 \left(\phi^\rho \phi_{\rho\sigma} \phi^\sigma\right)$,
${\cal L}_{A_4}/\sqrt{-g}
:=  A_4(X)  \phi^\sigma \phi_{\sigma\rho} \phi^{\rho\alpha} \phi_\alpha$, and 
$
{\cal L}_{A_5}/\sqrt{-g}
:= A_5(X)  (\phi^\rho \phi_{\rho\sigma} \phi^{\sigma})^2$.
The theory is also reflection-symmetric,
namely 
invariant under the transformation $\phi\to -\phi$. 
Varying the action with respect to $\phi$,
we obtain the EL equation of the scalar field 
\begin{eqnarray}
\label{sca_eq}
\nabla_\mu 
\left(
\frac{\partial {\cal L}}{\partial \phi_\mu}
\right)
-\nabla_\mu\nabla_\nu
\left(
\frac{\partial {\cal L}}{\partial \phi_{(\mu\nu)}}
\right)
=
\frac{\partial {\cal L}}{\partial \phi}.
\end{eqnarray}
In shift-symmetric theories,
i.e., $\partial {\cal L}/\partial \phi=0$,
and Eq. \eqref{sca_eq} can be rewritten 
into the form of the conservation law $\nabla_\mu J^\mu=0$,
where
\begin{eqnarray}
\label{noether}
J^\mu
:=
\frac{1}{\sqrt{-g}}
\left[
\frac{\partial {\cal L}}{\partial \phi_\mu}
-\nabla_\nu
\left(
\frac{\partial {\cal L}}{\partial \phi_{(\mu\nu)}}
\right)
\right],
\end{eqnarray}
corresponds to the Noether current associated with the shift symmetry.
Following Eq. \eqref{decomp},
the Noether current can be written as 
\begin{eqnarray}
J^\mu
=J^\mu_F
+\sum_{i=1}^5J_{A_i}^\mu,
\end{eqnarray}
where
\begin{eqnarray}
J^\mu_F
&:=&
2
\left(
F_{0X}+F_{2X}R
\right)
\phi^\mu,
\\
J_{A_1}^\mu
&:=&
2A_{1X} \phi^\mu \phi^{\alpha\beta}\phi_{\alpha\beta}
-2 A_{1X} X_\nu \phi^{\mu\nu}
-2A_1 \phi^{\mu\nu}{}_\nu,
\\
J_{A_2}^\mu
&:=&
2A_{2X} \phi^\mu (\Box\phi)^2
-2 A_{2X} X^\mu \Box\phi
-2A_2 \nabla^\mu (\Box\phi),
\\
J_{A_3}^\mu
&:=&
2A_{3X} \phi^\mu \Box\phi \left(\phi^\rho \phi_{\rho\sigma} \phi^\sigma\right)
+2A_3 (\Box\phi )\phi^\mu{}_\sigma \phi^\sigma
-A_{3X} X^\mu  \phi^\rho \phi_{\rho\sigma} \phi^\sigma
-A_3 \nabla^\mu
 \left(\phi^\rho \phi_{\rho\sigma} \phi^\sigma\right) 
\nonumber\\
&-&
A_{3X} X_\nu (\Box\phi)  \phi^\mu \phi^\nu
-A_3 \nabla_\nu (\Box\phi) \phi^\mu \phi^\nu
-A_3 \Box\phi \phi^\mu{}_\nu \phi^\nu
-A_3(\Box\phi)^2 \phi^\mu,
\\
J_{A_4}^\mu
&:=&
  2A_{4X} \phi^\mu \phi^\sigma \phi_{\sigma\rho} \phi^{\rho\alpha}\phi_\alpha
+2A_4 \phi^\mu{}_\rho \phi^{\rho\alpha}\phi_\alpha
-A_{4X} X_\nu
\left(
 \phi^\mu \phi^{\nu\alpha}\phi_\alpha
+\phi^\nu \phi^{\mu\alpha}\phi_\alpha
\right)
\nonumber\\
&-&
A_4
\left(
  \phi^\mu{}_\nu  \phi^{\nu\alpha} \phi_\alpha
+ \phi^\mu  \phi^{\nu\alpha}{}_\nu \phi_\alpha
+ \phi^\mu \phi^{\nu\alpha} \phi_{\alpha\nu}
+\Box\phi \phi^{\mu\alpha}\phi_\alpha
+\phi^\nu \phi^{\mu\alpha}{}_\nu\phi_\alpha 
+\phi^\nu \phi^{\mu\alpha} \phi_{\alpha\nu}
\right),
\\
J_{A_5}^\mu
&:=&
2A_{5X}\phi^\mu
     \left(\phi^\rho\phi_{\rho\sigma} \phi^\sigma\right)^2
+4 A_5  \left(\phi^\rho\phi_{\rho\sigma} \phi^\sigma\right)
         \phi^\mu{}_\alpha \phi^\alpha
-2 A_{5X}X_\nu
     \left(\phi^\rho\phi_{\rho\sigma} \phi^\sigma\right)
 \phi^\mu \phi^\nu
\nonumber\\
&-&2 A_{5}
   \nabla_\nu \left(\phi^\rho\phi_{\rho\sigma} \phi^\sigma\right)
 \phi^\mu \phi^\nu
-2 A_5 \left(\phi^\rho\phi_{\rho\sigma} \phi^\sigma\right)
 \phi^\mu{}_\nu \phi^\nu
-2A_5 \left(\phi^\rho\phi_{\rho\sigma} \phi^\sigma\right) 
  \phi^\mu{} \Box \phi,
\end{eqnarray}
with $X_\nu:=\nabla_\nu X$ and $X^\nu:= g^{\mu\nu}X_\mu$.

\subsection{Static and spherically symmetric spacetime}

We assume the static and spherically symmetric spacetime metric 
\begin{eqnarray} 
\label{sss}
ds^2 = - A(r)dt^2 + \frac{dr^2}{B(r)}+r^2(d\theta^2+\sin^2\theta d\varphi^2),
\end{eqnarray}
and the scalar field which shares the same symmetry with the spacetime 
\begin{eqnarray}
\phi=\phi (r).
\end{eqnarray} 
In this ansatz $X=B\phi'^2$,
where a `prime' denotes a derivative with respect to $r$, e.g. $\phi':=\partial \phi/\partial r$.
We note that 
in contrast to the case of the Horndeski and GLPV theories \cite{Kobayashi:2014eva,Babichev:2017guv},
the Noether current in DHOST theories 
contains the second-order and third-order derivative terms of the scalar field,
$\phi'' (r)$ and $\phi'''(r)$.

The EL equation for the scalar field \eqref{sca_eq} is given by
$0=
\partial_\mu 
\left(
\sqrt{-g}J^\mu
\right)
\propto
\partial_r 
\left(
r^2
\sqrt{A/B}
J^r
\right)$,
which can be integrated as 
$r^2\sqrt{A/B} J^r =C_0$,
where 
$C_0$ is an integration constant.
We impose the regularity of the norm of the Noether current everywhere on and outside the event horizon
$J_\mu J^\mu=(1/B)(J^r)^2=C_0^2/(r^4 A)<\infty$.
The regularity of $J_\mu J^\mu$ at the event horizon ($A\to 0$ and $B\to 0$) imposes $C_0=0$,
and hence the EL equation for the scalar field reduces to $J^r=0$.
Then, 
since $J_\mu J^\mu=0$, it also vanishes at the cosmological horizon, if it exists.

On the other hand,
the EL equations for the metric functions $A$ and $B$,
denoted by
$E_A=0$ and $E_B=0$ in the rest respectively,
can be obtained
by substituting the ansatz \eqref{sss} into the Lagrangian density \eqref{qdaction}
and then varying with respect to these variables.
One might worry about the missing independent components of 
the metric equations,
since in Eq. \eqref{sss} we have already fixed the gauge \cite{Motohashi:2016prk}.
However, as long as we focus on the ansatz $\phi=\phi(r)$,
the EL equations for $\phi$, $A$, and $B$
obtained from the gauge-fixed ansatz Eq. \eqref{sss}
provide a closed set of the equations.

\subsection{The degenerate equations}

First, 
using $\phi'=\pm \sqrt{X/B}$,
the equations $E_A=E_B=J^r=0$ 
written in terms of $\phi'$, $\phi''$ and $\phi'''$
can be rewritten 
into those 
in terms of $X$, $X'$, and $X''$
as 
\begin{eqnarray}
\label{eJ}
0
&=&
J^r
= C_{JX} (A,B,X) X'' 
+ C_{JA} (A,B,X) A''
+ C_{JB} (A,A',B,X,X') B'
+ {\Delta J}^r (A,A',B,X,X'),
\\
\label{ea}
0
&=&
E_A
= C_{AX} (A,B,X) X'' 
+ C_{AA} (A,B,X) A''
+ C_{AB} (A,A',B,X,X') B'
+ {\Delta E}_A (A, A', B, X, X'),
\\
\label{eb}
0
&=&
E_B
= C_{BX} (A,B,X) X'' 
+ C_{BA} (A,B,X) A''
+ C_{BB} (A,A',B,X,X') B'
+ \Delta E_B (A,A',B,X,X'),
\end{eqnarray}
where 
$C_{PQ}$ (where the indices $P$ and $Q$ are $P=J,A,B$ and $Q=X,A,B$ respectively),
$\Delta J^r$, $\Delta E_A$, and $\Delta E_B$
are the functions of the given variables.
Here, we do not show their (extremely lengthy) explicit form.
We note that because of the reflection symmetry under $\phi\to -\phi$
without loss of generality we may choose $\phi'= \sqrt{X/B}$.
The coefficients for Eqs. \eqref{eJ} and \eqref{eb} 
satisfy  
\begin{eqnarray}
 \frac{ C_{JX} }{C_{BX}}
=\frac{ C_{JA} }{C_{BA}}
=\frac{ C_{JB} }{C_{BB}}.
\end{eqnarray}
By combining Eq. \eqref{eb} with Eq. \eqref{eJ},
the terms proportional to $X''$, $A''$, and $B'$ are automatically canceled,
and 
$B$ can be obtained by solving $C_{BX} \Delta J^r-C_{JX}\Delta E_B=0$
in terms of the other variables:
\begin{eqnarray}
\label{binv}
\frac{1}{B}
&=&\frac{1}{32A (r^2 F_0+2F_2) (F_2-A_1 X)}
\left[
 (8F_2+r (A_3 X+4F_{2X})X'
-2A_1 (rX'+4X)
\right]
\nonumber\\
&\times&
\left[
   8r (F_2-A_1 X)A'
+A \left(8F_2+3r(A_3 X+4F_{2X})X'-2A_1 (4X+3rX')\right)
\right].
\end{eqnarray}
If we obtain the solutions for $X$ and $A$,
$B$ can be determined via Eq. \eqref{binv}.

Next, 
by eliminating $B$ and $B'$ in Eqs. \eqref{eJ} and \eqref{ea} with Eq. \eqref{binv},
\begin{eqnarray}
\label{eJ3}
0&=&J^r
= D_{JX} (A,A',X,X') X'' 
+ D_{JA} (A,A',X,X') A''
+ {\bar J}^r (A,A',X,X'),
\\
0&=&
E_A
= D_{AX} (A,A',X,X') X'' 
+ D_{AA} (A,A',X',X') A''
+ {\bar E}_A (A,A',X,X'),
\end{eqnarray}
where 
$D_{PQ}$ (with $P=J,A$ and $Q=X,A$),
${\bar J}^r$, and ${\bar E}_A$
are functions of the given variables.
Here,
we find that under the degeneracy conditions \eqref{2n1}
the coefficients satisfy
\begin{eqnarray}
  D_{JX}  D_{AA}- D_{JA} D_{AX}=0,
\end{eqnarray}
and hence 
\begin{eqnarray}
\label{det}
\Delta E(A,A',X, X')
:= 
\frac{{\bar J}^r}{D_{JX}}
-\frac{\bar{E}_A}{D_{AX}}
=0,
\end{eqnarray}
which gives a constraint on $A$ and $X$.
After some manipulation, Eq. \eqref{det} reduces to 
\begin{eqnarray}
\label{constraint}
-A_3 (3r^2 F_0+4F_2)X
-4\left[
  F_2 (2X A_{1X}-r^2 F_{0X})
+ r^2 F_0 (XA_{1X}+2F_{2X})
\right]
+2A_1
\left(
r^2 F_0-2 X (r^2 F_{0X}+2F_{2X})
\right)
=0.
\nonumber\\
\end{eqnarray}
Eq. \eqref{constraint} provides the algebraic relation $X=X(r)$,
and then 
substituting it into $J^r=0$ and $E_A=0$ reduces to the degenerate second-order 
differential equation for $A$.

As reviewed in Appendix \ref{appa_1},
in the case of a constant $X$,
the above method correctly reproduces
the Schwarzschild-(anti-) de Sitter solution obtained in Refs. \cite{Motohashi:2019sen,Takahashi:2019oxz}.

\subsection{The cases of a nonconstant $X$}

We next investigate BH solutions with a nonconstant $X$.
In order to obtain them,
we need to specify the coupling functions $F_{0,2}(X)$ and $A_{1,3}(X)$
(and all the remaining functions can be found via the degeneracy conditions \eqref{2n1}).
Before going to the more concrete models,
we briefly discuss two special models which lead to
a constant $X$.

\subsubsection*{Special case 1: $A_1=0$ and $A_3=0$}

First, we consider the models with 
\begin{eqnarray}
A_1=0,
\qquad
A_3=0,
\end{eqnarray}
which satisfy 
both the constraints of $c_{gw}=c$ 
\cite{Monitor:2017mdv,Creminelli:2017sry,Ezquiaga:2017ekz,Baker:2017hug,Sakstein:2017xjx}
and 
no gravitons decay into dark energy \cite{Creminelli:2018xsv}.
Then, degeneracy conditions \eqref{2n1} provide
$A_4=6F_{2X}^2/F_2$
and
$A_5=0$.
Thus,  $\Delta E=0$ gives rise to $F_2 F_{0X}-2F_0 F_{2X}=0$.
Unless 
one chooses the coupling function given by
$F_0=c_0 F_2^2$ with $c_0$ being constant,
the solution to $F_2 F_{0X}-2F_0 F_{2X}=0$
is always given by a constant $X$.

\subsubsection*{Special case 2: $F_0=0$ and $A_3=0$}

The similar subclass of the Class Ia quadratic DHOST theories 
is given by 
\begin{eqnarray}
F_0=0,
\qquad
A_3=0,
\end{eqnarray}
which also has 
no graviton decay into dark energy \cite{Creminelli:2018xsv}.
The degeneracy conditions \eqref{2n1} become
$A_4=
-(4XA_1-3F_2) (A_1-2F_{2X})^2/
        (2 (F_2-XA_1)^2)$
and
$A_5=
A_1 (A_1-2F_{2X})^2
        /(2 (F_2-XA_1)^2)$.
Thus,  $\Delta E=0$ gives rise to $F_2 A_{1X}+A_1 F_{2X}=0$
whose solution is always given by a constant $X$,
unless one chooses $A_1=c_0 /F_2$.

\subsubsection*{Models in the Class Ia  DHOST theories}

In order to explore explicit BH solutions with a nonconstant $X$ 
in the Class Ia quadratic DHOST theories, 
we will discuss the following models:

\begin{itemize}

\item
Model A
\begin{eqnarray}
\label{modb}
F_0(X)=\eta_1 X^n,
\qquad 
F_2(X)=\zeta,
\qquad
A_1(X)= \frac{\beta_1\zeta}{X},
\qquad
A_3(X)=\frac{\gamma_1 \zeta}{X^2},
\end{eqnarray}
where $\eta_1$, $\beta_1$, and $\gamma_1$ are constants
and we will assume $n>0$.
In this and the other models,
$\zeta$ is the constant related to the Planck mass squared $\zeta=M_{\rm Pl}^2/2$,
and hence we assume that $\zeta>0$.
Degeneracy conditions \eqref{2n1} determine the remaining coupling functions as 
\begin{eqnarray}
A_4=-\frac{(-12\beta_1^2+16\beta_1^3-12\beta_1\gamma_1 +\gamma_1 (8+\gamma_1)) \zeta}
         {8(1-\beta_1)^2X^2},
\qquad
A_5=\frac{(2\beta_1-\gamma_1) (2\beta_1^2-4\gamma_1+3\beta_1\gamma_1)\zeta}
         {8(1-\beta_1)^2X^3},
\end{eqnarray}
These coupling functions are singular at $X=0$.

\item
Model B
\begin{eqnarray}
\label{modd}
F_0(X)=-
\Lambda,
\qquad 
F_2(X)=\zeta-\beta_2X^n,
\qquad
A_1(X)= 0,
\qquad
A_3(X)=\gamma_2 X^{n-2},
\end{eqnarray}
where $\Lambda$ and $\gamma_2$ are constants.
Eqs. \eqref{2n1} determine 
\begin{eqnarray}
A_4= 
\frac{-48n^2X^{2n} \beta_2^2+8n X^{2n} \beta_2\gamma_2
 +\gamma_2 X^n \left(X^n(-8\beta_2+\gamma_2)+8\zeta\right)
}
        {8X^2 (\beta_2 X^n-\zeta)}, 
\qquad
A_5=\frac{X^{2n-3}\gamma_2 (4n\beta_2-\gamma_2)} 
            {2(\beta_2 X^n-\zeta)}.
\end{eqnarray}
This model satisfies $c_{gw}=c$.

\item
Model C
\begin{eqnarray}
\label{mode}
F_0(X)=-\Lambda,
\qquad 
F_2(X)=\gamma_3 X^n,
\qquad
A_1(X)= \beta_3 X^{n-1},
\qquad
A_3(X)=0,
\end{eqnarray}
where $\gamma_3$ and $\beta_3$ are constants.
Eqs. \eqref{2n1} determine
\begin{eqnarray}
A_4= 
-\frac{X^{n-2} (4\beta_3-3\gamma_3)(\beta_3-2n\gamma_3)^2}
        {2(\beta_3-\gamma_3)^2},
\qquad
A_5
=\frac{X^{n-3} \beta_3 (\beta_3-2n\gamma_3)^2}
        {2(\beta_3-\gamma_3)^2}.
\end{eqnarray}
This model does not have a smooth limit to GR or classic scalar-tensor theories,
as there is no constant term $\zeta$ in $F_2$.

\end{itemize}

\section{Exact black hole solutions and stability against the odd-parity perturbations}
\label{sec3}

We now present BH solutions for Models A-C, given by Eqs. \eqref{modb}, \eqref{modd}, and \eqref{mode},
where the detailed derivation of the BH solutions is presented in Appendix \ref{appb}.
We also argue the stability against odd-parity perturbations.

\subsection{Exact black hole solutions}

\subsubsection{Model A}

We first consider Model A~\eqref{modb} with the assumptions that $n>0$, $\eta_1>0$, $\beta_1<0$, and  
\begin{eqnarray}
\label{pspace3}
4n(1-\beta_1)>3(\gamma_1-2\beta_1)>0.
\end{eqnarray}
We will require the absence of the deficit solid angle,
which is that the coefficient of the $dr^2$
term becomes unity in the large distance limit $r\to\infty$, so 
the 3-space metric reduces to that of an Euclid space $dr^2+r^2(d\theta^2+\sin^2\theta d\varphi^2)$,
and 
otherwise the solution possesses
a deficit solid angle.
This requires
$\gamma_1
=2\beta_1
+ \frac{4n}{3}\sqrt{\beta_1(\beta_1-1)}$,
and then obtain the BH solution
\begin{eqnarray}
\label{solb}
A(r)
=
r^{2\sqrt{\frac{\beta_1}{\beta_1-1}}}
B(r),
\qquad
B(r)
=
1
-2M 
r^{-1- 
     \sqrt{\frac{\beta_1}
            {\beta_1-1}}},\\
X= 
\left(
\frac{
4\zeta [-\beta_1 +\sqrt{\beta_1 (\beta_1-1)}]
}
{3r^2\eta_1}
\right)^{\frac{1}{n}}.
\end{eqnarray}
The Schwarzschild solution with $X=0$ can be obtained in the limit of $\beta_1\to 0$
while keeping $\eta_1\neq 0$.
See Appendix \ref{appb_1} for further details of the derivation.

\subsubsection{Model B}

We consider Model B~\eqref{modd}.
We impose the conditions
\begin{eqnarray}
\label{pspace5}
\beta_2>0,
\qquad
\gamma_2>0.
\end{eqnarray}
We also impose 
\begin{eqnarray}
\label{pspace5_2}
-(3\gamma_2-8n\beta_2)\Lambda>0.
\end{eqnarray}
For the absence of the deficit solid angle,
we impose  $\gamma_2=\frac{4n\beta_2}{\sqrt{3}}$,
and then find the BH solutions 
\begin{eqnarray}
\label{sold}
A(r)
=r^{-2+ 2\sqrt{3} }
B(r),
\qquad
B(r)
=
1
-\frac{2M }{r^{2+\sqrt{3}}}.
\end{eqnarray}
In the metric solution Eq. \eqref{sold},
the event horizon exists at $A=B=0$.
The solution to $X$ is given by 
\begin{eqnarray}
X
=
\left(
\frac{4\zeta+(2\sqrt{3}-3)r^2\Lambda}
      {4\beta_2}
\right)^{\frac{1}{n}}.
\end{eqnarray}
From Eq. \eqref{pspace5_2},
we find that 
$0<-(3\gamma_2-8n\beta_2)\Lambda=4n\beta_2(2-\sqrt{3})\Lambda$
and hence $\Lambda>0$.
The effect of the bare cosmological constant $\Lambda$
appears only in the sector of the scalar field.
See Appendix \ref{appb_2} for further details of the derivation.

\subsubsection{Model C}

Finally, we consider Model C~\eqref{mode}.
We impose the conditions
\begin{eqnarray}
\label{pspace6}
\Lambda<0,
\qquad
\frac{1}{2}>n>0,
\qquad
\gamma_3
>\frac{3-2n}{4n}\beta_3
>\beta_3
>0.
\end{eqnarray}
For the absence of the deficit solid angle,
we impose 
$\gamma_3=
\frac{\beta_3}{6n}
\left(
7n
+
\sqrt{27+n (-108+109n)}
\right)$,
and find the BH solution
\begin{eqnarray}
\label{sole}
A(r)
=r^{2\frac{-3+5n+ \sqrt{27+n(109n-108)}}{3(1-2n)}}
B(r),
\qquad
B(r)
=
1
-2M 
r^{-\frac{6-13n+ \sqrt{27+n(109n-108)}}
           {3(1-2n)}}.
\end{eqnarray}
The solution to $X$ is given by 
\begin{eqnarray}
X=
\left(
-
\frac{n (-6+13n+\sqrt{27+n(109n-108)})}
      {2(1-2n)(10n-3)\beta_3}
\Lambda r^2
\right)^{1/n},
\end{eqnarray}
where the combination inside the round bracket
is positive for $0<n<1/2$, $\beta_3>0$, and $\Lambda<0$.
In the metric solution Eq. \eqref{sole},
the event horizon exists at $A=B=0$.
For $n=3/7$, Eq. \eqref{sole} reduces to
\begin{eqnarray}
\label{sole2}
A(r)
=B(r),
\qquad
B(r)
=
1
-\frac{2M}{r^3},
\qquad
X=\left(
-\frac{\Lambda}{2\beta_3}
r^2
\right)^{1/n},
\end{eqnarray}
which describes the asymptotically flat BH solutions.
The solution has no Schwarzschild limit,
as Model C has no limit to GR and to the ordinary scalar-tensor theories.
See Appendix \ref{appb_3} for further details of the derivation.

\subsection{A short summary}

We have derived static and spherically-symmetric BH solutions
with a noncontant $X=X(r)$
in several models of the Class Ia quadratic DHOST theories.
There are several common and different features among these solutions, which will be summarized here.

First, 
in all Models A-C,
BH solutions obtained in this section
are neither Schwarzschild nor Schwarzschild (-anti)- de Sitter solutions.
The two metric variables $A$ and $B$ satisfy 
$A=r^{c'}B $ with $c'>0$ being a constant,
which ensures that $A$ and $B$ cross zero at the same position
and such a position corresponds to an event horizon.
In all the models,
$A$ and $B$ cross zero once and
hence the solution describes a BH spacetime.
It is clear that 
from Eq. \eqref{binv}
the ratio $A/B$ is a nontrivial function of $r$ 
for a nonconstant kinetic term $X=X(r)$.  
In other words, 
Schwarzschild and Schwarzschild (-anti)- de Sitter solutions
would generically contain a constant kinetic term $X=X_0$.
The measurements of any deviation from the exact Schwarzschild [(-anti)-de Sitter]
 spacetime geometry is very important to
discriminate these models from GR or ordinary scalar-tensor theories satisfying the no-hair theorem
\cite{Carter:1971zc,Chase,Bekenstein:1972ny,Bekenstein:1995un,Graham:2014mda,Graham:2014ina}.
Since all the solutions discussed in this section possess the scalar field with $\phi=\phi(r)$,
the character of $\partial_\mu \phi$ is spacelike.

Second,
the existence of the Schwarzschild limit depends on the model.
As in the case of Model C,
if the theory does not have a smooth limit to GR,
obviously,
the resultant BH solution does not have a smooth limit to the Schwarzschild solution.
However,
as in the case of Model B,
even if the theory appears to have a limit to GR,
the BH solution may not have a smooth limit to the Schwarzschild solution.
When the nonminimal derivative coupling parameter $\beta_2$
is taken to be zero,
$X$ blows up
and hence the decoupling limit between the scalar and gravitation
is ill-defined.
On the other hand, 
Model A has a smooth limit to the scalar-tensor theory with the kinetic term $X^n$
and to the Schwarzschild solution
in the limit of $\beta_1\to 0$ and $\gamma_1\to 0$,
as the higher-order derivative terms are smoothly decoupled. 

From the astrophysical viewpoint, 
it is also important to ask
whether Newtonian gravity is recovered in the limit of $r\to \infty$.
In all Models A-C,
the BH solutions are not asymptotically flat.
The only exceptional case is that of $n=3/7$ in Model C (see Eq.~\eqref{sole2}),
where the gravitational potential scales as $1/r^3$,
different from the Newtonian $1/r$.
To avoid the conflict with the weak gravity tests of GR,
it may be reasonable to assume that 
the scalar field is localized in the vicinity of a BH
and does not affect physics in the Solar System.
However, 
in Models A-C, except for the case of $n=3/7$ in Model C,
the leading-order term in the gravitational potential 
defined by $\psi = (1/2)\lim_{r\to \infty} (A-1)$ 
grows with $r^{c'}$ ($c'>0$ being a constant)
and the deviation from Newtonian gravity becomes rather significant
far away from the BH.
Since in Model A the Schwarzschild solution is recovered as $\beta_1\to 0-$,
measurements would significantly constrain $\beta_1$,
while Models B and C (with $n\neq 3/7$) should be excluded. 
In the case of  $n=3/7$ in Model C,
as $\psi \sim 1/r^3$,
the existence of the scalar field around a BH 
would not affect physics in the asymptotic region.

Although we do not show it in the main text, 
we could also find peculiar BH solutions with a nonconstant $X$ 
for specific models as  
\begin{itemize}

\item Model D
\begin{eqnarray}\label{eq:modelD}
&&
F_0(X)=\eta_4 X,
\quad 
F_2(X)=\zeta+ \beta_4 \sqrt{X},
\quad
A_1(X)= 2F_{2X},
\quad
A_3(X)=\frac{\zeta}{X^2},
\end{eqnarray}
and from Eq. \eqref{2n1}
$A_4=-\frac{\sqrt{X} \beta_4 +9\zeta}
              {8X^2}$
and 
$A_5=\frac{\sqrt{X} \beta_4 +4\zeta}{8X^3}$.

\item Model E
\begin{eqnarray}\label{eq:modelE}
&&
F_0(X)=\alpha_5 X^2,
\quad 
F_2(X)=\zeta,
\quad
A_1(X)= \gamma_5 X,
\quad
A_3(X)=0,
\end{eqnarray}
and from Eq. \eqref{2n1}
$A_4=
\frac{\gamma_5^2 X^2 (3\zeta-4\gamma_5 X^2)}
        {2 (\zeta-X^2\gamma_5)^2}$
and 
$A_5=\frac{X^3\gamma_5^3}
            {2(\zeta-X^2\gamma_5)^2}$.

\end{itemize}
The explicit solutions in these examples are shown in Appendices \ref{appb_4} and \ref{appb_5},
respectively.

\subsection{Stability against odd-parity perturbations}

Finally, 
we investigate the stability of the BH solutions with a nonconstant kinetic term $X$
obtained in this section against odd-parity perturbations.
Our analysis follows the criteria obtained in Ref.~\cite{Takahashi:2019oxz}
for generic subclasses of the Class Ia DHOST theories. 
We leave the stability analysis against the even-parity perturbations for the future work.

In the case of $\phi=\phi(r)$,
the absence of ghost and Laplacian instabilities for the modes with the multipole indices $\ell \geq 2$ 
requires the conditions
\begin{eqnarray}
\label{hyp}
	F_2>0,\qquad 
        F_2-XA_1>0. \label{stability2}
\end{eqnarray}
Using the ${\cal S}$-deformation method,
the mode stability against the odd-parity perturbations with $\ell \geq 2$
is ensured if there exists a function 
\begin{eqnarray}
\label{s}
{\cal S}
=
\frac{d}{dr_\ast}
\left[
\ln 
\left(
r\sqrt{F_2}
\right)
\right]
=\sqrt{AB}
\left(
\frac{1}{r}
+\frac{F_{2X}}{2F_2}
 \frac{dX}{dr}
\right),
\end{eqnarray}
which is finite at both the boundaries, namely the event horizon and the spatial infinity.
We note that 
dipole perturbations of $\ell =1$ are related to the slow-rotation of a BH~\cite{Takahashi:2019oxz}.

\subsubsection{Model A}

In the case of Model A \eqref{modb},
the conditions \eqref{hyp} reduce to  
\begin{eqnarray}
\zeta (1-\beta_1)>0,
\end{eqnarray}
which with $\zeta>0$ and $\beta_1<0$
can be always compatible with Eq. \eqref{pspace3}.

On the other hand,
since $F_{2X}=0$ in Eq. \eqref{s}, we obtain
${\cal S}
=\sqrt{AB}/r$, which is definitively regular at the event horizon
where $A=B=0$,
and also regular at the infinity $r\to \infty$
as ${\cal S}\simeq r^{\sqrt{\beta_1/(\beta_1-1)}-1}\to 0$
for $\beta_1<0$.
Thus, 
the BH solution \eqref{solb} is stable against odd-parity perturbations.

\subsubsection{Model B}

In the case of Model B \eqref{modd},
the absence of the ghost and Laplacian instabilities requires
\begin{eqnarray}
F_2=F_2-A_1X
=\frac{r^2\Lambda}{4}
  (3-2\sqrt{3})
<0.
\end{eqnarray}
Thus, 
the solution \eqref{sold} suffers from 
either ghost or Laplacian instabilities.

\subsubsection{Model C}

In the case of Model C \eqref{mode},
the absence of the ghost and Laplacian instabilities requires
\begin{eqnarray}
F_2
=\gamma_3 X^n
>0,
\qquad
F_2-A_1 X
= (\gamma_3-\beta_3)X^n
>0,
\end{eqnarray}
which is always compatible with Eq. \eqref{pspace6}.

On the other hand,
${\cal S}
=2\sqrt{AB}/r$
is definitively regular at the event horizon
where $A=B=0$,
and also regular at the infinity $r\to \infty$
as ${\cal S}\propto r^{(-6+11n+\sqrt{27+n(-108+109n)})/[3(1-2n)]}\to 0$
for $0<n<1/2$.
Thus, 
the BH solution \eqref{sole} is stable
against odd-parity perturbations.

Before closing this section,
let us briefly mention the issues of the even-parity perturbations and the strong coupling problem.
Although the analysis of them is left for future work,
we expect
that the BH solutions with $\phi=\phi(r)$ obtained in this section 
would suffer from a strong coupling problem in the scalar field perturbations.
Ref. \cite{Minamitsuji:2018vuw}
found that
in a class of the shift-symmetry breaking Horndeski theory with $c_g=c$,
a stealth Schwarzschild solution with $\phi=\phi(r)$ and $X=X(r)$
was obtained,
but 
also suffered from the strong coupling problem.
The problem may arise
because in the background of $\phi=\phi(r)$
the kinetic term for the perturbation vanishes 
at the level of the quadratic order of the action.
A similar discussion was given in Ref. \cite{BenAchour:2019fdf}.
This motivates us to consider the general ansatz $\phi=qt+\psi(r)$
with the timelike character of the scalar field $X<0$ in the next section.

\section{The case of the linearly time-dependent scalar field}
\label{sec5}

We turn to the case of the linearly time dependent scalar field 
\cite{Babichev:2013cya,Kobayashi:2014eva,Babichev:2017lmw,BenAchour:2018dap,Motohashi:2019sen,Takahashi:2019oxz,Minamitsuji:2019shy}.
\begin{eqnarray}
\phi=q t+ \phi (r).
\end{eqnarray}
For simplicity, in addition to degeneracy conditions \eqref{2n1},  we impose 
\begin{eqnarray}
\label{a1a2=0}
A_1=A_2=0,
\end{eqnarray}
which satisfies $c_{gw}=c$.

\subsection{Solving the equations of motion}

In order to keep the independent EL equations after varying the action \eqref{qdaction},
we start with the general gauge of the static and spherically symmetric spacetime metric \cite{Motohashi:2016prk,Motohashi:2019sen,Takahashi:2019oxz}
\begin{eqnarray}
\label{ssq}
ds^2 = - A(r)dt^2 + \frac{dr^2}{B(r)}+2D(r) dt dr+r^2(d\theta^2+\sin^2\theta d\varphi^2).
\end{eqnarray}
After varying with respect to the all variables, we set $D=0$.
The nontrivial components of the Noether current are  
$J^t$ and $J^r$,
and  the equation for $\phi$ is given by
$0=
\partial_\mu 
\left(
\sqrt{-g}J^\mu
\right)
=\sqrt{-g} \partial_t J^t
+\partial_r 
\left(
\sqrt{-g}J^r
\right)
=\partial_r 
\left(
\sqrt{-g}J^r
\right)$,
which can be integrated as $r^2\sqrt{A/B} J^r=C_0$,
where $C_0$ is an integration constant.
The regularity condition of $J_\mu J^\mu=-A (J^t)^2+ (J^r)^2/B=-A (J^t)^2+C_0^2/(r^4A) <\infty$
on and outside the event horizon
imposes $C_0=0$,
and hence the scalar field equation reduces to $J^r=0$.
We note that $J_\mu J^\mu=0$ also on the cosmological horizon, if it exists.
On the other hand,
deriving the EL equations for $D$, $E_D$, and setting $D=0$
gives $E_D= (r^2/\sqrt{AB}) q J^r$,
as shown in Ref.~\cite{Babichev:2015rva}.
Thus, 
for $q\neq 0$,
$E_D=0$ is equivalent to $J^r=0$.

In terms of $A$, $B$, and $\phi'=\pm \sqrt{q^2/(AB)+X/B}$,
where because of the reflection symmetry without loss of generality we choose $\phi'= \sqrt{q^2/(AB)+X/B}$,
one can rewrite $E_A=E_B=J^r=0$ 
as 
\begin{eqnarray}
\label{eJq}
0
&=&
J^r
= C^q_{JX} (A,B,X) X'' 
+ C^q_{JA} (A,B,X) A''
+ C^q_{JB} (A,A',B,X,X') B'
+ \Delta_q J^r (A,A',B,X,X'),
\\
\label{eaq}
0
&=&
E_A
= C^q_{AX} (A,B,X) X'' 
+ C^q_{AA} (A,B,X) A''
+ C^q_{AB} (A,A',B,X,X') B'
+ \Delta_q E_A (A, A', B, X, X'),
\\
\label{ebq}
0
&=&
E_B
= C^q_{BX} (A,B,X) X'' 
+ C^q_{BA} (A,B,X) A''
+ C^q_{BB} (A,A',B,X,X') B'
+ \Delta_q E_B (A,A',B,X,X'),
\end{eqnarray}
where 
$C^q_{PQ}$ (with $P=J,A,B$ and $Q=X,A,B$),
$\Delta_q J^r$, $\Delta_q E_A$, and $\Delta_q E_B$
are the functions of the given variables.
The coefficients for Eqs. \eqref{eJq} and \eqref{ebq} 
satisfy  
$ C^q_{JX} /C^q_{BX}=C^q_{JA} /C^q_{BA}=C^q_{JB} /C^q_{BB}$.
By combining Eq. \eqref{ebq} with Eq. \eqref{eJq},
the terms proportional to $X''$, $A''$, and $B'$ automatically cancel
and $B$ can be written in terms of other variables:
\begin{eqnarray}
\label{binvq}
\frac{1}{B}
&=&
\frac{8F_2+r (A_3 X+4F_{2X})X'}{32A (r^2 F_0+2F_2) F_2}
\left[
  4r \left(2F_{2}A'+ q^2 A_3 X'\right)
+ A \left(8F_2+3r (A_3X+4F_{2X})X'\right)
\right].
\end{eqnarray}
By substituting $B$ and $B'$ into 
Eqs. \eqref{eJq} and \eqref{eaq}
\begin{eqnarray}
\label{eJq2}
0&=&J^r
= D^q_{JX} (A,A',X,X') X'' 
+ D^q_{JA} (A,A',X,X') A''
+ {\bar J}_q^r (A,A',X,X'),
\\\label{eaq2}
0&=&
E_A
= D^q_{AX} (A,A',X,X') X'' 
+ D^q_{AA} (A,A',X',X') A''
+ {\bar E}^q_A (A,A',X,X'),
\end{eqnarray}
where 
$D^q_{PQ}$ (with $P=J,A$ and $Q=X,A$),
${\bar J}_q^r$, and ${\bar E}^q_A$
are functions of the given variables.
With the degeneracy conditions \eqref{2n1} (with $A_1=A_2=0$),
we find $D^q_{JX}  D^q_{AA}- D^q_{JA} D^q_{AX}=0$,
and hence 
\begin{eqnarray}
\label{detq}
\Delta_q E
:= 
\frac{{\bar J}_q^r}{D^q_{JX}}
-\frac{\bar{E}^q_A}{D^q_{AX}}
=0.
\end{eqnarray}
In Appendix \ref{appa_2}
we review the case of a constant kinetic term, 
which as expected gives the BH model ``Case 1-$\Lambda$'' in Ref. \cite{Motohashi:2019sen}.

In the case of $q\neq 0$, it is more 
difficult to obtain nontrivial BH solutions.
$\Delta_q E=0$ provides
\begin{eqnarray}
\label{und}
0&=&
r A_3^2 X X'
\left[
4F_2 (5q^2 +3AX)
+r^2F_0 (14q^2+9AX)
\right]
-16r^2 (F_2F_{0X}-2F_0F_{2X})
\left(
2rF_2A'+A (2F_2+3r F_{2X}X')
\right)
\nonumber\\
&+&
4A_3 
\left[
r^2F_0
\left(
  F_2 (4q^2+6AX+6rXA')
+5r (2q^2+3AX)F_{2X}X'
\right)
\right.
\nonumber\\
&&
\left.
+
F_2
\left(
 8F_2 (q^2+AX +rXA')
+
r
\left(
-r^2 (4q^2+3AX)F_{0X}
+4 (q^2+3AX)F_{2X}
\right)
X'
\right)
\right],
\end{eqnarray}
in which $A$ and $X$ (and their derivatives)
are mixed up.
In the limit of $q\to 0$,
Eq. \eqref{und} can be product separable and
reduce to 
$0=3r^2A_3F_0X
+4A_3F_2X
-4r^2F_2 F_{0X}
+8r^2 F_0 F_{2X}$,
which of course
agrees with Eq. \eqref{constraint} with $A_1=0$.

As reviewed in Appendix \ref{appa_2},
in the case of a constant $X$,
this method correctly reproduces
the Schwarzschild-(anti-) de Sitter solution obtained in Refs. \cite{Motohashi:2019sen,Takahashi:2019oxz}.

For BHs with a nonconstant $X$,
we consider the two examples

\begin{itemize}

\item Model A-q

The first model we consider is given by 
\begin{eqnarray}
\label{q_model}
F_0=-\Lambda,
\qquad
F_2=\zeta>0,
\qquad
A_3=
\frac{\gamma\zeta}{X^2}.
\end{eqnarray}

\item Model B-q

The second model is given by 
\begin{eqnarray}
\label{q_model2}
F_0=-\beta X,
\qquad
F_2=\zeta>0,
\qquad
A_3=
\frac{\gamma\zeta}{X^2}.
\end{eqnarray}

\end{itemize}

From the degeneracy conditions Eq. \eqref{2n1},
both Models A-q and B-q,
Eqs.
\eqref{q_model}
and 
\eqref{q_model2}
respectively,
provide 
$A_4
=-(\gamma(\gamma+8)\zeta)/(8X^2)$
and 
$A_5
=\gamma^2\zeta/(2X^3)$.

We note that if $F_0=0$ in both Models A-q and B-q,
the BH solution is given by the Schwarzschild metric 
$A=B=1-2M/r$ with $X=-q^2$.
Thus, adding a nonzero $F_0(X)$ modifies the asymptotic structure of the spacetime,
while the near-horizon geometry is close to the Schwarzschild solution.

\subsection{BH solutions with a nonconstant $X$}

\subsubsection{Model A-$q$}

First, we consider Model A-$q$ given by Eq.~\eqref{q_model}.
As shown in Appendix \ref{appc}, we find the exact BH solutions
expressed by
\begin{eqnarray}
\label{nonzeroq_sol}
X&=&
-
q^2
\left(1-\frac{3r^2\Lambda}{4\zeta}\right)^{-\frac{4}{4-3\gamma}},
\\
\label{nonzeroq_sol2}
A
&=&
-\frac{1}
      {3r(3\gamma-4)}
\left(
1-\frac{3r^2\Lambda}{4\zeta}
\right)^{\frac{3\gamma}{8-6\gamma}}
\left\{
 6M (-4+3\gamma)
+4r (2-5\gamma)
 {}_2F_1
\left[
\frac{1}{2},
-\frac{8-3\gamma}{8-6\gamma},
\frac{3}{2};
\frac{3r^2\Lambda}{4\zeta}
\right]
\right.
\nonumber\\
&&
\left.
+2r (2+5\gamma)
 {}_2F_1
\left[
\frac{1}{2},
-\frac{3\gamma}{8-6\gamma},
\frac{3}{2};
\frac{3r^2\Lambda}{4\zeta}
\right]
+\gamma r
 {}_2F_1
\left[
\frac{1}{2},
\frac{3}{2}
+\frac{2}{-4+3\gamma},
\frac{3}{2};
\frac{3r^2\Lambda}{4\zeta}
\right]
\right\}.\\
\label{nonzeroq_sol3}
\frac{B}{A}&=&
\frac{(4-3\gamma)^2\zeta^2}
       {\left( (4-3\gamma) \zeta+3r^2 (\gamma-1)\Lambda\right)^2}
\left(
1-\frac{3r^2\Lambda}{4\zeta}
\right)^{3+\frac{4}{-4+3\gamma}}.
\end{eqnarray}
From Eqs. \eqref{nonzeroq_sol}-\eqref{nonzeroq_sol3},
$X<0$ and the character of $\partial_\mu\phi$ is timelike. 
For $\Lambda>0$ and $\gamma<1$ or $\gamma>4/3$
(or for $\Lambda<0$ and $1<\gamma<4/3$),
there is a curvature singularity at the finite radius $r=r_{s}= \sqrt{(4-3\gamma)\zeta/ [3(1-\gamma)\Lambda]}$,
where both $R$ and $R^{\mu\nu\alpha\beta}R_{\mu\nu\alpha\beta}$ blow up.
In the case of $\gamma= 0$,
we recover the Schwarzschild-(anti-) de Sitter solutions
$A=B=1-(2M)/r-(\Lambda r^2)/(6\zeta)$.
We note that 
in the limit of $\gamma\to 0$, $q=0$, $\psi'=0$, and hence $X=0$.

\begin{figure}[h]
\unitlength=1.1mm
\begin{center}
  \includegraphics[height=5.1cm,angle=0]{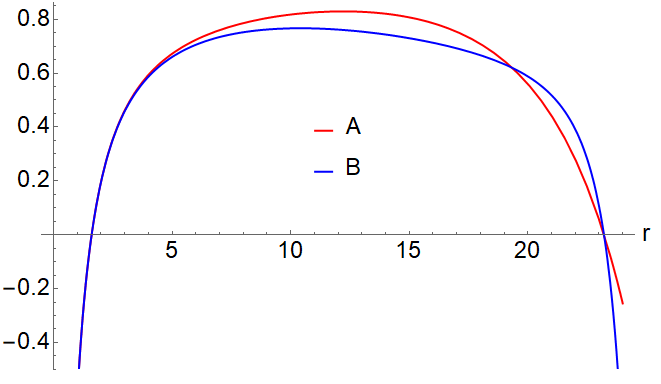} 
  \includegraphics[height=5.1cm,angle=0]{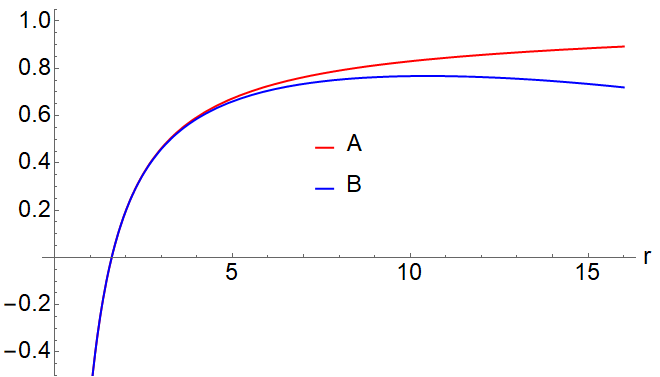}
\caption{
The metric functions $A$ and $B$ 
of the solution 
\eqref{nonzeroq_sol}
are shown 
by the red and blue curves respectively.
The left and right panels 
correspond to 
$\Lambda=10^{-3}$ and $\Lambda=-10^{-3}$,
respectively. 
The other parameters are
chosen as $\zeta=1$, $\gamma=2.0$, $M=0.8$, and $q=1.0$.
}
  \label{figsol}
\end{center}
\end{figure}

In the limit of $\Lambda=0$,
the BH solution coincides with the Schwarzschild solution
and the solution in the presence of a small $\Lambda$ is
given by 
\begin{eqnarray}
A
&=&
1-\frac{2M}{r}
 +\frac{r \left(8r -3 (9M+r)\gamma\right)}{12 (3\gamma-4)} 
   \frac{\Lambda}{\zeta}
+{\cal O} (\Lambda^2),
\qquad
B=
1-\frac{2M}{r}
 +\frac{r \left(9M\gamma+4r (2-3\gamma)\right)}{12 (3\gamma-4)} 
   \frac{\Lambda}{\zeta}
+{\cal O} (\Lambda^2), 
\nonumber\\
X
&=&
q^2\left[
-1
+\frac{3 r^2}
        {3\gamma-4}
 \frac{\Lambda}{\zeta}
+{\cal O} (\Lambda^2)
\right].
\end{eqnarray}
We note that 
for 
$\Lambda=0$
the solution exactly corresponds to the stealth Schwarzschild solution
with $X=-q^2$
and is consistent with the conditions for ``Case 1'' in Ref. \cite{Motohashi:2019sen},
and for $r\ll \sqrt{ |\zeta/\Lambda|}$,
the solution can be well approximated by the Schwarzschild solution.
Thus, on distance scales much shorter than $\sqrt{ |\zeta/\Lambda|}$
the gravitational force law can be approximated by the Newtonian one.
By choosing $M$ appropriately, 
no singularity appears 
between the two event and cosmological horizons.
In Fig. \ref{figsol},
the metric functions $A$ and $B$ 
of the solution Eqs. \eqref{nonzeroq_sol2} and \eqref{nonzeroq_sol3}
are shown 
by the red and blue curves, respectively.
Here, 
the left and right panels 
correspond to
 $\Lambda=10^{-3}$ and $\Lambda=-10^{-3}$,
respectively. 
The other parameters are
chosen to be $\zeta=1$, $\gamma=2.0$, $M=0.8$, and $q=1.0$.
For $\Lambda<0$,
for an appropriate choice of the parameters,
$A$ and $B$ cross zero once at the same position,
and hence the metric solution Eqs. \eqref{nonzeroq_sol2}-\eqref{nonzeroq_sol3}
represents a BH spacetime.
At an intermediate radius,
$B$ monotonically increases,
while $A$ starts to decrease but never crosses zero again.  
For $\Lambda>0$,
for an appropriate choice of parameters,
$A$ and $B$ cross zero twice at the same positions,
which may be identified
as the event and cosmological horizons, respectively.

\subsubsection{Model B-$q$}

Second, we consider Model B-$q$ given by Eq.~\eqref{q_model2}.
The difference from Model A-$q$ \eqref{q_model}
is that $F_0$ is replaced by $-\beta X$.
In this case, we cannot find an analytic solution
and instead need to solve numerically. 
Following the same procedure,
we obtain the set of equations to be integrated numerically
\begin{eqnarray}
X'
&=&
-\frac{8X}{r\gamma}
\left[
-4q^2\gamma \zeta
+r^2\beta (-4+3\gamma) X^2 (A+rA')
+2\gamma X \left(-2\zeta A+r (q^2r\beta-2\zeta A')\right)
\right]
\nonumber\\
&\times&
\left[
-20q^2 \gamma\zeta 
+2\left(q^2 r^2\beta (-8+7\gamma) -6r\zeta A\right)X
+3r^2 \beta (-4+3\gamma)AX^2
\right]^{-1},
\nonumber\\
A'
&=&
-\frac{2}{rX}
\left[
8q^2\zeta^2\gamma^2
+4\gamma\zeta 
 \left(
 q^2r^2\beta (4-5\gamma)
+2\gamma \zeta A
 \right)
X
+2r^2\beta\gamma
\left(
q^2r^2\beta (-8+5\gamma)
+3 (4-3\gamma)\zeta A
\right)
X^2
+r^4\beta^2(4-3\gamma)^2 A X^3
\right]
\nonumber\\
&\times&
\left[
 16\gamma^2\zeta^2
-24r^2 \beta (-2+\gamma)\gamma \zeta X
+r^4\beta^2 (32-36\gamma+9\gamma^2)X^2
\right]^{-1}.
\end{eqnarray}
We exclude $\gamma=0$,
which provides the Schwarzschild solution with $q=0$.

\begin{figure}[h]
\unitlength=1.1mm
\begin{center}
    \includegraphics[height=5.25cm,angle=0]{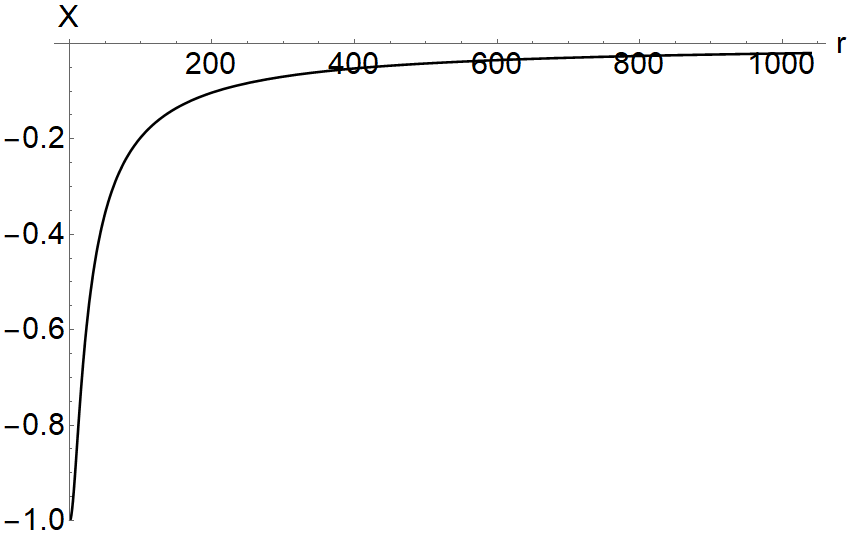} 
  \includegraphics[height=5.25cm,angle=0]{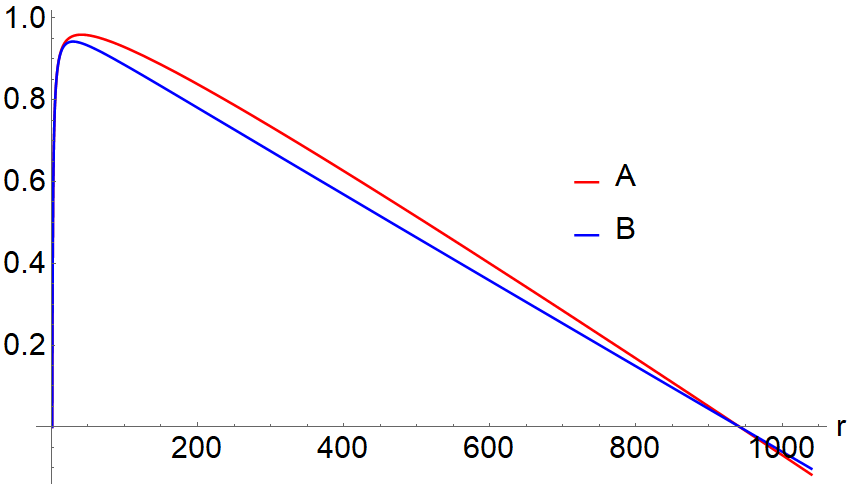}
\caption{
The kinetic term $X$ (the left panel)
and 
the metric functions $A$ and $B$ (the right panel)
are shown
as the functions of $r$.
In the right panel
$A$ and $B$ are shown by the red and blue curves respectively.
Here, 
we set the position of the event horizon at $r=1.0$
and $\zeta=1.0$, $q=1.0$, $\beta=-10^{-4}$, and $\gamma=0.05$,
and $X(r=1.0)=-1.0$ as the boundary condition.
}
  \label{figsol2}
\end{center}
\end{figure}

In Fig. \ref{figsol2},
for Model B-q \eqref{q_model2}
the kinetic term $X$ (the left panel)
and 
the metric functions $A$ and $B$ (the right panel)
are shown
as the functions of $r$.
In the right panel
$A$ and $B$ are shown by the red and blue curves respectively.
We set the position of the event horizon at $r=1.0$
and $\zeta=1.0$, $q=1.0$, $\beta=-10^{-4}$, and $\gamma=0.05$,
and $X(r=1.0)=-1.0$ as the boundary condition.
Throughout the domain of $r$,
$X<0$ and hence the character of $\partial_\mu\phi$ is timelike.
We also find
that $A$ and $B$ cross zero twice at $r=1.0$ and $r\approx 942.0$,
where the latter is interpreted as the cosmological horizon.
We note that 
in contrast to Model A-q \eqref{q_model} with $\Lambda>0$,
no singularity appears outside the cosmological horizon.

\begin{figure}[h]
\unitlength=1.1mm
\begin{center}
    \includegraphics[height=5.2cm,angle=0]{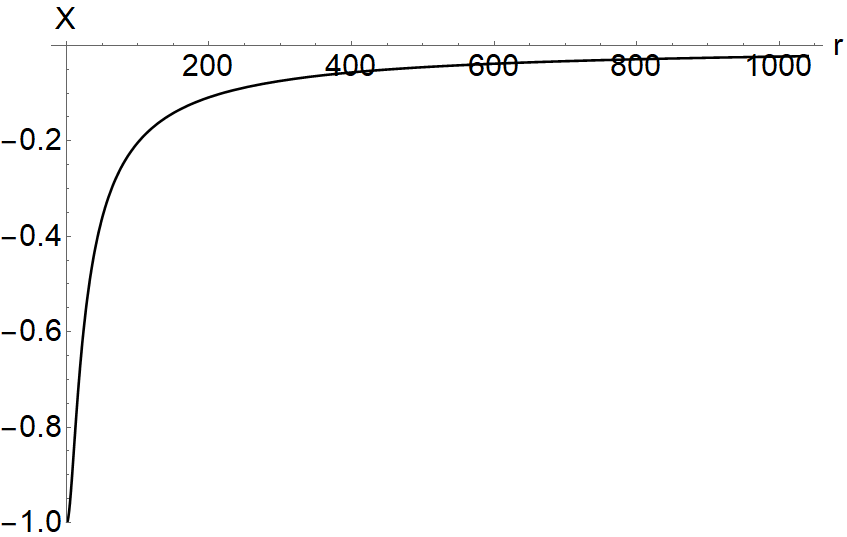}
    \includegraphics[height=5.2cm,angle=0]{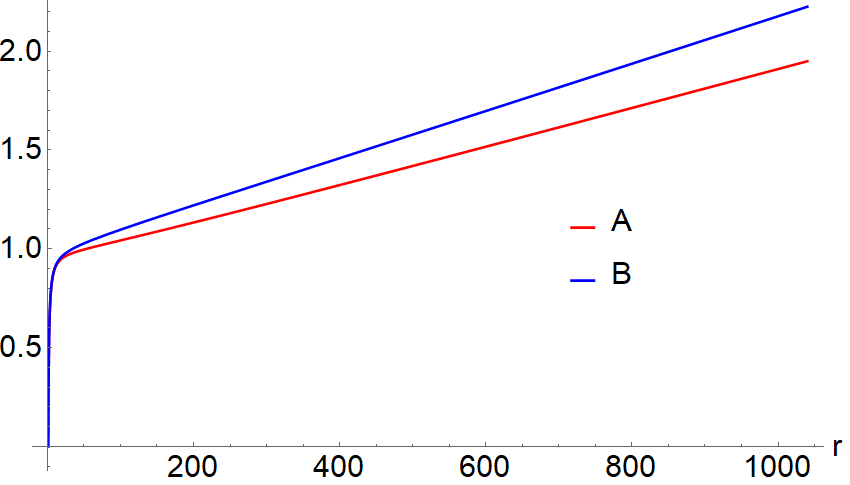}
\caption{
The kinetic term $X$ (the left panel)
and 
the metric functions $A$ and $B$ (the right panel)
are shown
as the functions of $r$.
In the right panel
$A$ and $B$ are shown by the red and blue curves, respectively.
Here, 
we set the position of the event horizon at $r=1.0$
and $\zeta=1.0$, $q=1.0$, $\beta=10^{-4}$, and $\gamma=-0.05$,
and $X(r=1.0)=-1.0$ as the boundary condition.
}
  \label{figsol3}
\end{center}
\end{figure}

In Fig. \ref{figsol3},
for Model B-q \eqref{q_model2}
the same plots are shown 
for $\zeta=1.0$, $q=1.0$, $\beta=10^{-4}$, and $\gamma=-0.05$,
and $X(r=1.0)=-1.0$ as the boundary condition.
Throughout the domain of $r$,
$X<0$ and hence the character of $\partial_\mu\phi$ is timelike.
We also find
that $A$ and $B$ cross zero only once at $r=1.0$ and monotonically increase.
As $\beta$ approaches zero,
$A$ and $B$ approach that of the Schwarzschild solution.

\subsection{Stability against odd-parity perturbations}

If the background scalar field is a function of the radial coordinate,  $\phi=qt+ \psi(r)$,
the absence of ghost and Laplacian instabilities for the modes with the multipole indices $\ell \geq 2$ 
requires the conditions
\begin{eqnarray}
\label{hyp2q}
	F_2>0,\qquad 
        F_2-XA_1>0, \qquad
        A_1\leq 0.
\label{stability2q}
\end{eqnarray}
Using the ${\cal S}$-deformation method,
the mode stability against the odd-parity perturbations with $\ell \geq 2$
is ensured 
the function
\begin{eqnarray}
\label{sq}
{\cal S}
=
\frac{d}{dr_\ast}
\left[
\ln 
\left(
r
\sqrt{
\frac{A{\cal F}{\cal G} +B{\cal J}^2}
       {A{\cal G}}
}
\right)
\right],
\end{eqnarray}
where the functions ${\cal F}$, ${\cal G}$, ${\cal H}$, and ${\cal J}$
are defined in Eq.~(36) of Ref.  \cite{Takahashi:2019oxz},
is finite at both the boundaries, namely both the event horizon and the spatial infinity
(or the cosmological horizon, if exists).

In Model A-q \eqref{q_model} with Eq.~\eqref{a1a2=0},
${\cal F}={\cal G}={\cal H}=2F_2=2\zeta>0$ and ${\cal J}=0$,
and hence the conditions Eq. \eqref{hyp2q} are automatically satisfied.
On the other hand,
${\cal S}=\sqrt{AB}/r$,
which always vanishes at the event horizon.
At spatial infinity, ${\cal S}$ always decay for $\gamma>4/3$
and the mode stability can be ensured for this case.
For $\gamma<4/3$
the above definition of ${\cal S}$ 
blows up at the infinity.
However, 
since there might be another choice of ${\cal S}$
which satisfy the desired properties,
this does not necessarily mean the instability.

Before closing this section,
let us also comment on the strong coupling problem.
In Ref. \cite{deRham:2019gha},
it was shown 
that the Schwarzschild -de Sitter BH solutions
with a constant kinetic term $X=X_0$
suffer the strong coupling problem. 
In Ref. \cite{Motohashi:2019ymr},
it was argued
that
in the context of the effective field theory
the strong coupling problem 
would be avoided by detuning of the degeneracy conditions \eqref{2n1}
for the Case 1 solutions with $X=-q^2$
at least in the cosmological asymptotic region,
where the apparent Ostrogradski ghost would appear
above the cutoff scale of the effective field theory.
It  has not been explicitly investigated  
whether the strong coupling problem exist
for the BH solutions with a nonconstant kinetic term.
In Models A-q and B-q,
the Case-1 Schwarzschild solution with $X=-q^2$
is recovered in the limits of $\Lambda\to 0$ and $\beta\to 0$,
respectively,
but
for $\Lambda \neq 0$ and $\beta\neq 0$
the deviation from the constant $X=-q^2$ solution
becomes more significant in the large distance regions,
as $X$ goes to $0$ in the limit of $r\to \infty$.
An explicit analysis about the even-parity perturbations will be necessary 
to clarify this issue.
If there is still the strong coupling problem,
it is also interesting to see
whether the cure argued in Ref. \cite{Motohashi:2019ymr}
could work or not.

\section{Conclusions}
\label{sec6}

We investigated static and spherically symmetric BH solutions 
with a nonconstant kinetic term of the scalar field 
in shift-symmetric Class Ia quadratic DHOST theories.
We used the method developed in Refs. \cite{Kobayashi:2018xvr,Takahashi:2019oxz}.
Because of the properties of the DHOST theories,
the highest-order derivative terms
of the EL equations in static and spherically symmetric backgrounds  
are degenerate,
and
combining them gives rise to a constraint relation.
Using this constraint relation,
the higher-order differential equations can be rewritten
in terms of the second-order differential equations.

In order to find analytic BH solutions with a nonconstant $X$,
we had to choose the particular form of the coupling functions. 
We found Models A, B, and C
given by Eqs. \eqref{modb}, \eqref{modd}, and \eqref{mode},
which could provide the analytic BH solutions
whose metrics are given by Eqs.~\eqref{solb}, \eqref{sold}, and \eqref{sole},
respectively.
These solutions have metrics
which are neither Schwarzschild nor Schwarzschild - (anti-) de Sitter solutions,
but possess only one BH event horizon.

We have also investigated the stability of
these BH solutions against the odd-parity perturbations.
Our analysis employed
the conditions for the absence of the ghost and Laplacian instabilities,
and those for the mode stability, 
which were obtained in Ref. \cite{Takahashi:2019oxz}.
We have shown 
both that the BH solutions in Models A and C
are free from ghost and Laplacian instabilities,
and also mode stable,
since the function ${\cal S}$ given by Eq.~\eqref{s}
was regular at both the event horizon and the spatial infinity
(or the cosmological horizon, if it exists).
However,
BH solutions in Model B suffer from ghost or Laplacian instabilities.

In \S~\ref{sec5},
we investigated BH solutions with a nonconstant kinetic term 
for a more general ansatz of the scalar field $\phi=qt+\psi (r)$ with a constant $q$.
We have clarified why it is more difficult  to find exact BH solutions compared to
$q=0$.
We considered Models A-q \eqref{q_model} and B-q \eqref{q_model2} with $A_1=A_2=0$.
In these examples, 
adding nontrivial functions of $F_0(X)$ modified the asymptotic structure of the spacetime,
while the near horizon geometry is close to that of the Schwarzschild BH.

Several issues are left for future work.
First, the stability of the BH solutions obtained in this paper
against the even-parity perturbations
and the possibilities of  the strong coupling problems \cite{deRham:2019gha}
could be investigated.
For coupling functions other than those studied in this paper,
one may still be able to construct new BH solutions with a nonconstant kinetic term.
Finally, 
construction of rotating BH solutions with a nonconstant kinetic term
will also be important topics for the future.

\acknowledgments{
M.M.~was supported by the research grant under the Decree-Law 57/2016 of August 29 (Portugal) through the Funda\c{c}\~{a}o para a Ci\^encia e a Tecnologia.
J.E. is grateful for the support recieved from CENTRA.
}

\appendix

\section{BH solutions with a constant $X$}
\label{appa}

We briefly review the case of the constant kinetic term $X=X_0$.

\subsection{The case of $\phi=\phi(r)$}
\label{appa_1}

For $\phi=\phi(r)$,
the equations for $B$ and $\Delta E$ reduce 
to
\begin{eqnarray}
\label{const_b}
B^{-1}
=\frac{ 2(F_2-X_0 A_1) (A+rA')}
        {A (r^2 F_0+2F_2) },
\end{eqnarray}
and 
\begin{eqnarray}
\Delta E
&=&\frac{32 (X_0 A_1-F_2) (A+rA')^2}
        {r^2(r^2 F_0+2F_2) (-2A+rA')^2 (-2A_1 +X_0 A_3 +4F_{2X})^2}
\left[
  X_0 A_3 (3r^2 F_0+4F_2)
+4r^2 X_0 F_0 A_{1X}
+8X_0 F_2 A_{1X}
\right.
\nonumber\\
&&
\left.
-
4r^2 F_2F_{0X}
+8r^2 F_0F_{2X}
+A_1
(-2r^2F_0+4r^2 X_0 F_{0X}+8X_0F_{2X})
\right],
\end{eqnarray}
where $F_{0,2}(X)$, $A_{0,1,2,3} (X)$, and their derivatives are evaluated at $X=X_0$. 
$\Delta E=0$,
together with the existence of the solution for $B$,
requires 
\begin{eqnarray}
\label{const}
&& X_0 A_3 (3r^2 F_0+4F_2)
+4r^2 X_0 F_0 A_{1X}
+8X_0 F_2 A_{1X}
\nonumber\\
&&
-
4r^2 F_2F_{0X}
+8r^2 F_0F_{2X}
+A_1
(-2r^2F_0+4r^2 X_0 F_{0X}+8X_0F_{2X})
=0.
\end{eqnarray}
Eq. \eqref{const} is satisfied for all $r$,
if the coefficients of the $r^2$ and $r^0$ terms separately vanish,
namely,   
\begin{eqnarray}
\label{const_cond}
A_3
&=&
\frac{  8F_0 F_2 A_{1X}
+2A_1^2 (F_0-2X_0F_{0X})
+A_1 (-4X_0 F_0 A_{1X}+4F_2 F_{0X})}{F_0(3X_0 A_1-4F_2)},
\nonumber\\
F_{2X}
&=&
\frac
{F_2
\left[
X_0F_0 A_{1X}
+2F_2 F_{0X}
+A_1 (F_0 -2X_0 F_{0X})
\right]
}
{F_0(-3X_0A_1+4F_2)}.
\end{eqnarray}
Substituting Eq. \eqref{const_cond} into $J^r=0$ and $E_A=0$
provides the degenerate equation
$r\left(r^2 F_0+2F_2\right)A''
+4F_2A'
-2rF_0A=0$,
which can be solved as
\begin{eqnarray}
\label{a0}
A(r)
=
C_1 
\left[
2F_2
+\frac{F_0}{3}r^2
-\frac{4M(F_2-X_0 A_1)}{r}
\right],
\end{eqnarray}
where $C_1$ and $M$ are integration constants,
corresponding to the degrees of freedom for the rescaling of time and the mass of the BH. 
Substituting into Eq. \eqref{const_b}
\begin{eqnarray}
\label{b0}
B(r)
=
-\frac{\Lambda}{3}r^2
+\kappa
-\frac{2M}{r},
\end{eqnarray}
where we have defined 
$\Lambda:= -F_0/[2(F_2-X_0 A_1)]$,
and
$\kappa:=F_2/(F_2-X_0 A_1)$.
By choosing $C_1$ appropriately,  it is possible to make $A=B$.
This solution described the Schwarzschild-(anti-) de Sitter-type solution
with the effective cosmological constant $\Lambda$ 
\cite{Motohashi:2019sen,Takahashi:2019oxz}.
The deficit solid angle is absent ($\kappa=1$) in the case of $A_1=0$,
namely in the case 
that the difference between $c_{gw}$, the speed of GWs and $c$, the speed of light 
is absent, i.e. $c_{gw}=c$.

\subsection{The case of $\phi=qt+\psi(r)$}
\label{appa_2}

For $\phi=qt+\psi(r)$,
the equation for $B$ becomes
\begin{eqnarray}
\label{const_b2q}
B^{-1}
=\frac{2F_2 (A+rA')}
        {A (r^2 F_0+2F_2) },
\end{eqnarray}
and \eqref{detq} becomes
\begin{eqnarray}
\Delta_q E
\propto
A_3 
\left[
4F_2 (q^2+X_0 A+r X_0 A')
+r^2 F_0(2q^2 +3X_0 A+3rX_0 A')
\right]
-
4r^2 (A+rA')
(F_2F_{0X}-2F_0 F_{2X}).
\end{eqnarray} 
$\Delta E_q=0$ sets
\begin{eqnarray}
\label{con1}
A'(r)
&=&
\left[
-A_3 
 \left(
   r^2 (2q^2 +3X_0A)F_0
+4 (q^2+X_0 A)F_2 
 \right)
+4r^2 A
(F_2F_{0X}-2F_0 F_{2X})
\right]
\nonumber\\
&\times&
\left[
  rX_0 A_3 (3r^2 F_0+ 4F_2)
-4r^3 (F_2F_{0X}-2F_0 F_{2X})
\right]^{-1}.
\end{eqnarray}
Then,
$J^r=0$ and $E_A=0$ yield the degenerate condition
\begin{eqnarray}
\label{con2}
F_{2X}
=-\frac{3}{8}X_0 A_3 +\frac{F_2F_{0X}}{2F_0}.
\end{eqnarray}
Substituting Eq. \eqref{con2} into Eq. \eqref{con1},
\begin{eqnarray}
 A(r)
=-\frac{X_0}{q^2}
\left(
\frac{F_0}{6F_2}r^2
+1-\frac{2M}{r}
\right).
\end{eqnarray}
From Eq. \eqref{const_b2q},
we find
\begin{eqnarray}
 B(r)
=
\frac{F_0}{6F_2}r^2
+1-\frac{2M}{r}.
\end{eqnarray}
If we choose $X_0=-q^2$,
the solution can be rewritten into the form of 
the standard Schwarzschild -(anti-) de Sitter solution:
$A(r)= B(r)
=[F_0/(6F_2)]r^2+1-2M/r$,
which corresponds to ``Case 1-$\Lambda$'' in Ref. \cite{Motohashi:2019sen}.

\section{BH solutions with a nonconstant $X$ for $\phi=\phi(r)$}
\label{appb}

In this appendix, we show the derivation process of BH solutions presented in \S~\ref{sec3}.

\subsection{Model A}
\label{appb_1}

We first consider Model A~\eqref{modb}.
From Eqs. \eqref{det} and \eqref{binv}, 
\begin{eqnarray}
X=
\left(\frac{4(\gamma_1-2\beta_1)\zeta}
          {r^{2}\eta_1(4n(1-\beta_1)-3(\gamma_1-2\beta_1))}
\right)^{1/n}.
\end{eqnarray}
Hereafter, for $n>0$,
we impose the conditions Eq. \eqref{pspace3},
and assume that $\beta_1<0$.
Substituting it into Eq. \eqref{binv},
\begin{eqnarray}
\label{nonc_b3}
B=\frac{16n^2(1-\beta_1)A}
          {\left(4n(1-\beta_1)-3(\gamma_1-2\beta_1)\right)
           \left[
             (4n(1-\beta_1)-3(\gamma_1-2\beta_1))A
           +4nr(1-\beta_1)A'
           \right]}.
\end{eqnarray}
Then, 
the degenerate equations $J^r=0$ and $E_A=0$ yield the second-order 
differential equation for $A$,
whose solution is given by
\begin{eqnarray}
A=
C_1
r^{-\left(4n(1-\beta_1) +6\beta_1-3\gamma_1 \right)/[4n(1-\beta_1)]} 
+
C_2
r^{-3(2\beta_1-\gamma_1)/[2n(1-\beta_1)]}, 
\end{eqnarray}
with $C_1$ and $C_2$ being integration constants.
Substituting into Eq. \eqref{nonc_b3},
\begin{eqnarray}
\label{eqb_b}
B=
\frac{16n^2 (1-\beta_1)}
      {\left(4n(1-\beta_1)-6\beta_1+3\gamma_1\right)
       \left(4n(1-\beta_1)+6\beta_1-3\gamma_1\right)}
\left(
1
+\frac{C_1}{C_2}
r^{-\left(4n(1-\beta_1)-6\beta_1+3\gamma_1\right)/[4n(1-\beta_1)]}
\right).
\end{eqnarray}
When the conditions Eq. \eqref{pspace3} are satisfied, 
the second term in Eq. \eqref{eqb_b}
becomes negligible in the large distance limit $r\to \infty$ and hence the spatial  metric approaches 
\begin{eqnarray}
\frac {\left(4n(1-\beta_1)-6\beta_1+3\gamma_1\right)
        \left(4n(1-\beta_1)+6\beta_1-3\gamma_1\right)}
       {16n^2 (1-\beta_1)}
     dr^2
     +r^2(d\theta^2+\sin^2\theta d\varphi^2).
\end{eqnarray}
If the coefficient of the $dr^2$
term becomes unity,
the 3-space metric reduces to that of an Euclid space $dr^2+r^2(d\theta^2+\sin^2\theta d\varphi^2)$,
and 
otherwise the solution possesses
a deficit solid angle.

In order to avoid the appearance of 
the deficit solid angle,
we also impose 
\begin{eqnarray}
\frac{16n^2 (1-\beta_1)}
      {\left(4n(1-\beta_1)-6\beta_1+3\gamma_1\right)
       \left(4n(1-\beta_1)+6\beta_1-3\gamma_1\right)}
=1,
\end{eqnarray} 
which yields the two roots
\begin{eqnarray}
\gamma_1
=\gamma_{1,\pm}
=2\beta_1
\pm \frac{4n}{3}\sqrt{\beta_1(\beta_1-1)}.
\end{eqnarray}
Since $\gamma_{1,\pm}-2\beta_1=\pm 4n\sqrt{\beta_1(\beta_1-1)}/3$,
only the plus branch is consistent with Eq. \eqref{pspace3}. 
For $\gamma_1=\gamma_{1,+}$,
choosing $C_1=-2M$ and $C_2=1$ with $M$ being a constant, 
\begin{eqnarray}
\label{solb_app}
A(r)
=
r^{2\sqrt{\frac{\beta_1}{\beta_1-1}}}
B(r),
\qquad
B(r)
=
1
-2M 
r^{-1- 
     \sqrt{\frac{\beta_1}
            {\beta_1-1}}}.
\end{eqnarray}
The solution of $X$ is given by 
\begin{eqnarray}
X= 
\left(
\frac{
4\zeta [-\beta_1 +\sqrt{\beta_1 (\beta_1-1)}]
}
{3r^2\eta_1}
\right)^{\frac{1}{n}}.
\end{eqnarray}
The Schwarzschild solution can be obtained in the limit of $\beta_1\to 0$.
This is consistent with the fact
that as $\beta_1\to 0$ and then $\gamma_{1,+}\to 0$
Model A reduces to the scalar-tensor theory with the kinetic term $X^n$.

\subsection{Model B}
\label{appb_2}

We consider Model B~\eqref{modd}.
From Eqs. \eqref{det} and \eqref{binv},
\begin{eqnarray}
X=
\left(\frac{-r^2(3\gamma_2-8n\beta_2)\Lambda+4\gamma_2\zeta}
          {4\beta_2\gamma_2}
\right)^{1/n}.
\end{eqnarray}
From $n>0$,
we impose the conditions
\begin{eqnarray}
\label{pspace5}
\beta_2>0,
\qquad
\gamma_2>0.
\end{eqnarray}
We also impose 
$-r^2(3\gamma_2-8n\beta_2)\Lambda+4\gamma_2\zeta>0$
for any $r$ and hence Eq. \eqref{pspace5_2}.
Then, 
the degenerate equations $J^r=0$ and $E_A=0$ 
yield the second-order differential equation for $A$,
whose solution is given by 
\begin{eqnarray}
A=r^{-4+\frac{3\gamma_2}{4n\beta_2}}
\left[
C_1
+
C_2
r^{2+\frac{3\gamma_2}{4n\beta_2}}
\right],
\end{eqnarray}
where $C_1$ and $C_2$ are integration constants.
Substituting into the equation for $B$, we find
\begin{eqnarray}
B
&=&
\frac{16n^2\beta_2^2}
       {64n^2\beta_2^2-9\gamma_2^2}
\left[
1+
\frac{C_1}{C_2}
r^{-2-\frac{3\gamma_2}{4n\beta_2}}
\right].
\end{eqnarray}
For the absence of the deficit solid angle,
we impose 
$16n^2\beta_2^2/(64n^2\beta_2^2-9\gamma_2^2)=1$,
which 
with Eq. \eqref{pspace5}
can be solved as $\gamma_2=\frac{4n\beta_2}{\sqrt{3}}$.
For this choice of $\beta_4$, 
by setting $C_1=-2M$ and $C_2=1$,
we find 
\begin{eqnarray}
\label{sold_app}
A(r)
=r^{-2+ 2\sqrt{3} }
B(r),
\qquad
B(r)
=
1
-\frac{2M }{r^{2+\sqrt{3}}}.
\end{eqnarray}
In the metric solution Eq. \eqref{sold},
the event horizon exists at $A=B=0$.
The solution to $X$ is given by 
\begin{eqnarray}
X
=
\left(
\frac{4\zeta-(3-2\sqrt{3})r^2\Lambda}
      {4\beta_2}
\right)^{\frac{1}{n}}.
\end{eqnarray}

\subsection{Model C}
\label{appb_3}

We then consider Model C~\eqref{mode}.
From Eqs. \eqref{det} and \eqref{binv}, 
\begin{eqnarray}
X=
\left(
-
    \frac{(2n-3)\beta_3+4n \gamma_3}
          {4(1-2n)\beta_3\gamma_3}
\Lambda r^2
\right)^{1/n}.
\end{eqnarray}
For $n>0$,
we impose the conditions \eqref{pspace6}.
The solution for $B$ is given by 
\begin{eqnarray}
B
=\frac{4n^2 (\beta_3-\gamma_3)\gamma_3 A}
        {\left[(2n-3)\beta_3+4n\gamma_3\right]
         \left[
          \left((2n+3)\beta_3-8n\gamma_3\right)A
         +2n r(\beta_3-\gamma_3)A'
         \right]}.
\end{eqnarray}
The degenerate equations $J^r=0$ and $E_A=0$ 
yield the second-order differential equation for $A$,
whose solution is given by 
\begin{eqnarray}
A=
C_1
r^{  \frac{(-3+4n)\beta_3+2n\gamma_3}
            {n(\beta_3-\gamma_3)}
}
+
C_2
r^{-\frac{3\beta_3+2n\beta_3-8n\gamma_3}
           {2n(\beta_3-\gamma_3)}
   },
\end{eqnarray}
where $C_1$ and $C_2$ are integration constants.
Substituting into the equation for $B$, we find
\begin{eqnarray}
B
&=&
\frac{4n^2\gamma_3(\beta_3-\gamma_3)}
       {\left(
         (10n-3)\beta_3-4n\gamma_3\right)
       \left(
         (-3+2n)\beta_3+4n\gamma_3\right)}
\left[
1+
\frac{C_2}{C_1}
r^{\frac{3\beta_3-10n\beta_3+4n\gamma_3}
          {2n(\beta_3-\gamma_3)}}
\right].
\end{eqnarray}
For the absence of the deficit solid angle,
we impose 
$4n^2\gamma_3(\beta_3-\gamma_3)/[
       \left(
         (10n-3)\beta_3-4n\gamma_3\right)
       \left(
         (-3+2n)\beta_3+4n\gamma_3\right)]
=1$,
which 
with Eq. \eqref{pspace5}
can be solved as 
\begin{eqnarray}
\gamma_3=
\frac{\beta_3}{6n}
\left(
7n
\pm 
\sqrt{27+n (-108+109n)}
\right).
\end{eqnarray}
We note
that 
$27+n (-108+109n)>0$ for $0<n<1/2$,
and only the plus branch satisfies Eq. \eqref{pspace6}
and hence the minus branch is excluded in the rest of this subsection.
For this choice of $\gamma_3$, 
by setting $C_2=-2M$ and $C_1=1$,
we find 
\begin{eqnarray}
\label{sole_app}
A(r)
=r^{2\frac{-3+5n+ \sqrt{27+n(-108+109n)}}{3(1-2n)}}
B(r),
\qquad
B(r)
=
1
-2M 
r^{-\frac{6-13n+ \sqrt{27+n(-108+109n)}}
           {3(1-2n)}}.
\end{eqnarray}
In the metric solution Eq. \eqref{sole},
the event horizon exists at $A=B=0$.
For $n=3/7$, Eq. \eqref{sole} reduces to
\begin{eqnarray}
\label{sole3}
A(r)
=B(r),
\qquad
B(r)
=
1
-\frac{2M}{r^3},
\end{eqnarray}
which describes the asymptotically flat BH solutions.
The solution to $X$ is given by 
\begin{eqnarray}
X=
\left(
-
\frac{n (-6+13n+\sqrt{27+n(-108+109n)})}
      {2(1-2n)(-3+10n)\beta_3}
\Lambda r^2
\right)^{1/n},
\end{eqnarray}
where the combination inside the round bracket
is positive for $0<n<1/2$, $\beta_3>0$, and $\Lambda<0$.
The solution has no Schwarzschild limit,
as Model C has no limit to GR and to the classic scalar-tensor theories.

\subsection{Model D}
\label{appb_4}

Although we do not consider them in details in the text, we
will introduce two other models here for reference.

For Model D~\eqref{eq:modelD},
by substituting into Eqs. \eqref{det} and \eqref{binv}, 
we obtain
\begin{eqnarray}
\label{dea}
\Delta E
&=&
\frac{(-4\zeta +r^2 \eta_4 X) (8X+rX')
      \left(8r X A'+A(8X+3rX')\right)^2}
      {16r^2X^2 (2\zeta+2\beta_4 \sqrt{X}+r^2\eta_4 X)(rA'-2A)^2}
=0,
\\
\label{ba}
B
&=&
\frac{32 A X^2 (2\zeta+2\beta_4 \sqrt{X}+r^2\eta_4 X)}{\zeta (8X+rX')\left(8r X A'+A(8X+3rX')\right)}.
\end{eqnarray}
In order for Eq. \eqref{dea} to be satisfied while allowing a solution of $B$ to exist, we impose
\begin{eqnarray}
X=\frac{4\zeta}{r^2\eta_1}.
\end{eqnarray}
Substituting this into Eq. \eqref{ba},
we obtain 
\begin{eqnarray}
\label{nonc_b}
B=\frac{16A}
          {3\zeta (A+4rA')}
\left(
3\zeta+\frac{2\beta_4}{r} \sqrt{\frac{\zeta}{\eta_4}}
\right).
\end{eqnarray}
Then,
the degenerate equations $J^r=0$ and $E_A=0$ 
yield the second-order differential equation for $A$
\begin{eqnarray}
2r
\left[
4r \left(3r\zeta +2\beta_4 \sqrt{\frac{\zeta}{\eta_4}}\right)A''
+\left(-3r\zeta +6\beta_4   \sqrt{\frac{\zeta}{\eta_4}}\right)A'
\right]
-\left(9r\zeta +2\beta_4 \sqrt{\frac{\zeta}{\eta_4}}\right)A
=0,
\end{eqnarray}
Finding the solution of $A$ and substituting it back into Eq. \eqref{nonc_b},
we obtain 
\begin{eqnarray}
\label{solD_app}
B=\frac{A}{r^{3/2}}
  = \frac{16}{7}
  +\frac{32\beta_4}{9\zeta r}\sqrt{\frac{\zeta}{\eta_4}}
   -\frac{2M}{ r^{7/4}}.
\end{eqnarray}
The solution is neither Schwarzschild nor Schwarzschild (anti-) de Sitter.
Since $A$ and $B$ cross zero once and at the same place,
the solution describes a BH spacetime.
As $r\to \infty$, $B\to 16/7$ and hence
the three-dimensional space has a deficit solid angle.

\subsection{Model E}
\label{appb_5}

Another example  we consider is Model E~\eqref{eq:modelE}.
From Eqs. \eqref{det} and \eqref{binv}, 
\begin{eqnarray}
X=
\frac{2}{r}
\left(
\frac{(r^2\alpha_5-\gamma_5)\zeta}
      {5\alpha_5\gamma_5}\right)^{1/2}.
\end{eqnarray} 
For $n>0$,
we impose the conditions
\begin{eqnarray}
\label{pspace4}
\zeta>0, \qquad
\alpha_5>0,
\qquad
\gamma_5>0,
\end{eqnarray}
and the point where $r^2\alpha_5-\gamma_5=0$
is located inside the event horizon (see below).
Substituting it into Eq. \eqref{binv},
\begin{eqnarray}
\label{nonc_b4}
B=
\frac{r^2 \alpha_5 (2r^2\alpha_5+3\gamma_5) (r^2\alpha_5+4\gamma_5)A}
{\gamma_5 (r^2\alpha_5+3\gamma_5)
  \left[
   (r^2\alpha_5+\gamma_5)A+r (r^2\alpha_5+4\gamma_5)A'
  \right]}.
\end{eqnarray}
The degenerate equations $J^r=0$ and $E_A=0$ 
yield the second-order differential equation for $A$,
whose solution is given by 
\begin{eqnarray}
A
&=&
\frac{C_2}{21\sqrt{r} (r^2\alpha_5+4\gamma_5)^{3/4}}
\left[
7 
\left(
2r^4\alpha_5^2
-19 r^2\alpha_5 \gamma_5
-108 \gamma_5^2
+3 C_1 r^{1/4} (r^2\alpha_5+4\gamma_5)^{3/8}
\right)
\right.
\nonumber \\
&&
\left.
+
80\times 2^{1/4} \alpha_5 \gamma_5
r^2
  \left(4+\frac{\alpha_5 r^2}{\gamma_5}\right)^{3/8}
{}_2F_1
\left[
\frac{3}{8},
\frac{7}{8},
\frac{15}{8};
-\frac{\alpha_5}{4\gamma_5}r^2
\right]
\right].
\end{eqnarray}
with $C_1$ and $C_2$ being integration constants,
and 
${}_2F_1$ represents the hypergeometric function.
Substituting into Eq. \eqref{nonc_b4},
\begin{eqnarray}
\label{solc_app}
B
&=&
\frac{r^2\alpha_5}{21\gamma_5 (r^2\alpha_5+3\gamma_5)^2}
\left[7 \left(
 2r^4\alpha_5^2
-19 r^2\alpha_5 \gamma_5
-108 \gamma_5^2
+3C_1 r^{1/4} (r^2\alpha_5+4\gamma_5)^{3/8}
\right)
\right.
\nonumber \\
&&
\left.
+
80\times 2^{1/4} \alpha_5 \gamma_5 r^2
  \left(4+\frac{\alpha_5 r^2}{\gamma_5}\right)^{3/8}
{}_2F_1
\left[\frac{3}{8},
\frac{7}{8},
\frac{15}{8};
-\frac{\alpha_5}{4\gamma_5}r^2
\right]\right].
\end{eqnarray}
By choosing $C_2=\frac{\alpha_5}{\gamma_5}$,
\begin{eqnarray}
A
=
\frac{ \left(\alpha_5 r^2 +3\gamma_5\right)^2}
       {r^{5/2}\left(\alpha_5 r^2 +4\gamma_5\right)^{3/4}}
B.
\end{eqnarray}
$A$ and $B$ cross zero once at the same place and monotonically increase, and hence
the solution describes a BH spacetime with an event horizon.

\section{BH solutions with a nonconstant $X$ for $\phi=qt+\psi(r)$}
\label{appc}

In this appendix, we show the derivation process of BH solutions in Models A-q \eqref{q_model} and B-q \eqref{q_model2}.

\subsection{Model A-q}
\label{appc_1}

Eqs. \eqref{binvq} and \eqref{detq} can then be rewritten as 
\begin{eqnarray}
\frac{1}{B}
&=&
\frac{
 \zeta \left(8X+\gamma r X'\right)
\left[
8X^2 (A+rA')
+r\gamma \left(4q^2+3AX\right) X'
\right]
}
 {32 (2\zeta - r^2\Lambda)A X^3},
\nonumber\\
\Delta_q E
&=&
-
A^2 (8X+r\gamma X')^2
\left(
8X^2 (A+rA')
+r\gamma (4q^2+3AX)X'
\right)
\nonumber\\
&
\times
&
\left[
8 X 
\left(
2q^2 (2\zeta-r^2\Lambda)
+(4\zeta-3r^2\Lambda) X(A+rA')
\right)
+\gamma r
\left(
2q^2 (10\zeta-7r^2\Lambda)
+3 (4\zeta-3r^2\Lambda)
 A X
\right)
X'
\right]
\nonumber\\
&\times&
\left[
16\gamma r^2
(2\zeta-r^2\Lambda)
X
(4q^2+ 2A X-rXA')^2
\left(
 4q^4\gamma
+ X
\left(
 2q^2\gamma A
-q^2  r\gamma A'
+A^2 (8X+r\gamma X')
\right)
\right)
\right]^{-1}.
\end{eqnarray}
The equation $\Delta_q E=0$ gives
\begin{eqnarray}
X'
&=&
-
\frac{
8X
\left[
2q^2 (2\zeta -r^2\Lambda)
+(4\zeta-3r^2\Lambda)(AX
+r XA')
\right]}
{
r\gamma
\left(
 2q^2 (10\zeta-7r^2\Lambda)
+3(4\zeta-3r^2\Lambda)AX
\right)},
\end{eqnarray}
and therefore eliminating $X'$
\begin{eqnarray}
\label{appnonzeroq_sol0A}
B=
-
\frac{
AX
\left[
  2q^2 (10\zeta-7r^2\Lambda)
+ 3(4\zeta -3r^2\Lambda) AX
\right]^2
}
{
4q^2 \zeta (4\zeta-3r^2\Lambda)
\left(
4q^2+2AX-rXA'
\right)^2
}.
\end{eqnarray}
The degenerate equations \eqref{eJq2}
and \eqref{eaq2} reduce to
an equation for $A$: 
\begin{eqnarray}
A'
&=&
\left[
4q^2 
\left(
-4 (-4+3\gamma)\zeta^2
-10r^2 (2-3\gamma)\Lambda\zeta
+3r^4 (2-5\gamma)\Lambda^2
\right)
-2
(4\zeta-3r^2\Lambda)
\left(
(-8+6\gamma)\zeta
+3r^2 (2-3\gamma)\Lambda
\right)
AX
\right]
\nonumber\\
&\times&
\left[
r(-4+3\gamma)
 (4\zeta-3r^2\Lambda)^2
X
\right]^{-1}.
\end{eqnarray}
We note that 
our analysis excludes the case of $\gamma=4/3$.
We find the exact BH solutions
expressed by the following nonconstant kinetic term and the metric functions  
\begin{eqnarray}
\label{appnonzeroq_sol}
X&=&
-
q^2
\left(1-\frac{3r^2\Lambda}{4\zeta}\right)^{-\frac{4}{4-3\gamma}},
\nonumber\\
A
&=&
-\frac{1}
      {3r(3\gamma-4)}
\left(
1-\frac{3r^2\Lambda}{4\zeta}
\right)^{\frac{3\gamma}{8-6\gamma}}
\left\{
 6M (-4+3\gamma)
+4r (2-5\gamma)\,\,
 {}_2F_1
\left[
\frac{1}{2},
-\frac{8-3\gamma}{8-6\gamma},
\frac{3}{2};
\frac{3r^2\Lambda}{4\zeta}
\right]
\right.
\nonumber\\
&&
\left.
+2r (2+5\gamma)\,\,
 {}_2F_1
\left[
\frac{1}{2},
-\frac{3\gamma}{8-6\gamma},
\frac{3}{2};
\frac{3r^2\Lambda}{4\zeta}
\right]
+\gamma r\,\,
 {}_2F_1
\left[
\frac{1}{2},
\frac{3}{2}
+\frac{2}{-4+3\gamma},
\frac{3}{2};
\frac{3r^2\Lambda}{4\zeta}
\right]
\right\}.
\end{eqnarray}
$B$ can be calculated by obtained $A$ and $X$ into Eq. \eqref{appnonzeroq_sol0A}, namely
\begin{eqnarray}
\frac{B}{A}=
\frac{(4-3\gamma)^2\zeta^2}
       {\left( (4-3\gamma) \zeta+3r^2 (\gamma-1)\Lambda\right)^2}
\left(
1-\frac{3r^2\Lambda}{4\zeta}
\right)^{3+\frac{4}{3\gamma-4}}.
\end{eqnarray}

\subsection{Model B-q}
\label{appc_2}

Eqs. \eqref{binvq} and \eqref{detq} can then be rewritten as 
\begin{eqnarray}
\frac{1}{B}
&=&
-
\left[
\zeta (8 X+r\gamma X')
\left(
4r (2X^2A'+q^2\gamma X')
+AX (8X+3r\gamma X')
\right)
\right]
\nonumber\\
&\times&
\left[
32 A X^3 (2\zeta-r^2\beta X)
\right]^{-1},
\nonumber\\
\Delta_q E
&=&
A^2 (8X+r\gamma X')^2
\left(
4r (2X^2 A'+q^2\gamma X')
+AX (8X+3r\gamma X')
\right)
\nonumber\\
&\times&
\left[
\gamma  X^2 \left(16 r \left(\beta  q^2 r-2 \zeta 
   A'\right)+A \left(3 \beta  (3 \gamma -4) r^3 X'-32
   \zeta \right)\right)+8 \beta  (3 \gamma -4) r^2 X^3 \left(rA'+A\right)
\right.
\nonumber\\
&&
\left.
+2 \gamma  X \left(X' \left(\beta  (7
   \gamma -8) q^2 r^3-6 \gamma  \zeta  r A\right)-16 \zeta 
   q^2\right)-20 \gamma ^2 \zeta  q^2 r X'
\right]
\nonumber\\
&\times&
\left[
16 \gamma ^2 r^2 X \left(\beta  r^2 X-2 \zeta \right)
   \left(-r X A'+2 A(r) X+4 q^2\right)^2 
\right.
\nonumber\\
&&
\left.
\times
\left(\gamma 
   q^2 r X A'-2 \gamma  q^2 A X-A^2 X
   \left(\gamma  r X'+8 X\right)-4 \gamma  q^4\right)
\right]^{-1}.
\end{eqnarray}
The equation $\Delta_q E=0$ gives
\begin{eqnarray}
\label{eqx}
X'
&=&
-\frac{8X}{r\gamma}
\left[
-4q^2\gamma \zeta
+r^2\beta (-4+3\gamma) X^2 (A+rA')
+2\gamma X \left(-2\zeta A+r (q^2r\beta-2\zeta A')\right)
\right]
\nonumber\\
&\times&
\left[
-20q^2 \gamma\zeta 
+2\left(q^2 r^2\beta (-8+7\gamma) -6r\zeta A\right)X
+3r^2 \beta (-4+3\gamma)AX^2
\right]^{-1},
\end{eqnarray}
and  eliminating $X'$
\begin{eqnarray}
\label{appnonzeroq_sol0}
B&=&
-
\left[
A X \left(-2 X \left(\beta  (7 \gamma -8) q^2 r^2-6 A \gamma 
   \zeta \right)+3 A \beta  (4-3 \gamma ) r^2 X^2+20 \gamma  \zeta
    q^2\right)^2
\right]
\nonumber\\
&\times&
\left[
4 \gamma  \zeta  q^2 \left(4 q^2+2 A X-r X A'\right)^2 \left(4
   \gamma  \zeta +\beta  (4-3 \gamma ) r^2 X\right)
\right]^{-1}.
\end{eqnarray}
The degenerate equations \eqref{eJq2}
and \eqref{eaq2} reduce to
an equation for $A$: 
\begin{eqnarray}
\label{eqa}
A'
&=&
-\frac{2}{rX}
\left[
8q^2\zeta^2\gamma^2
+4\gamma\zeta 
 \left(
 q^2r^2\beta (4-5\gamma)
+2\gamma \zeta A
 \right)
X
+2r^2\beta\gamma
\left(
q^2r^2\beta (5-8\gamma)
+3 (4-3\gamma)\zeta A
\right)
X^2
+r^4\beta^2(4-3\gamma)^2 A X^3
\right]
\nonumber\\
&\times&
\left[
 16\gamma^2\zeta^2
-24r^2 \beta (\gamma-2)\gamma \zeta X
+r^4\beta^2 (32-36\gamma+9\gamma^2)X^2
\right]^{-1}.
\end{eqnarray}
The combined equations \eqref{eqx} and \eqref{eqa}
are solved numerically.

In the vicinity of the BH horizon $r=r_h$, the solution for $A$, $B$, and $X$ can be decomposed as
\begin{eqnarray}
A&=&a_1 (r-r_h)+ {\cal O} \left[(r-r_h)^2\right],
\nonumber \\
B&=&b_1 (r-r_h)+ {\cal O} \left[(r-r_h)^2\right],
\nonumber \\
X&=&x_0 + x_1 (r-r_h)+ {\cal O} \left[(r-r_h)^2\right],
\end{eqnarray} 
where
\begin{eqnarray}
a_1
&=&-\left[4 \gamma  q^2 \left(4 \gamma  \zeta ^2+\beta ^2 (5 \gamma -8) r_h^4
   x_0^2+2 \beta  (4-5 \gamma ) \zeta  r_h^2 x_0\right)\right]
\nonumber\\
&\times&
\left[
r_h x_0 \left(16 \gamma^2 \zeta ^2+\beta ^2 \left(9 \gamma ^2-36
   \gamma +32\right) r_h^4 x_0^2-24 \beta  (\gamma -2) \gamma  \zeta 
   r_h^2 x_0\right)
\right]^{-1},
\nonumber\\
b_1
&=&
-\frac{\left(\beta  (3 \gamma -8) r_h^2 x_0-4 \gamma  \zeta \right)
   \left(4 \gamma  \zeta ^2+\beta ^2 (5 \gamma -8) r_h^4 x_0^2+2
   \beta  (4-5 \gamma ) \zeta  r_h^2 x_0\right)}
         {16 r_h  \zeta (r_h^2 x_0\beta (\gamma-2)-\gamma\zeta)^2},
\nonumber\\
x_1
&=&
\frac{8r_h x_0^2 \beta}
       {r_h^2 x_0 \beta (3\gamma-8)-4\gamma\zeta},
\end{eqnarray}
which are used for the boundary conditions for the numerical integration.
In the limit of $\beta=0$, $x_1=0$,
$a_1=-q^2 /(r_h x_0)$,
and 
$b_1=1/r_h$,
which represents the Schwarzschild spacetime with $X=x_0$.

\bibliography{ref-BH}

\end{document}